\documentclass[a4paper,accepted=2024-10-14]{quantumarticle}
\pdfoutput=1

\usepackage[utf8]{inputenc}
\usepackage[T1]{fontenc}
\usepackage[colorlinks]{hyperref}
\usepackage[normalem]{ulem}
\usepackage{soul}
\usepackage{comment}

\usepackage[numbers,sort&compress]{natbib}

\usepackage{amsthm,amsfonts,amsmath,amssymb}
\usepackage{accents}
\usepackage{mathtools}
\usepackage{bbm}
\usepackage{bm}
\usepackage{upgreek} 
\usepackage{isomath}

\usepackage{graphicx}
\usepackage{tikz}
\usetikzlibrary{arrows}

\newtheorem{theorem}{Theorem}

\newtheorem{definition}[theorem]{Definition}

\def\bra#1{\mathinner{\langle{#1}|}}
\def\ket#1{\mathinner{|{#1}\rangle}}

\def\ketbra#1#2{\mathinner{|{#1}\rangle\!\langle{#2}|}}

\def\BraVert{\egroup\,\mid@vertical\,\bgroup}

\def\dket#1{\mathinner{|{#1}\rangle\!\rangle}}
\def\dketbra#1#2{\mathinner{|{#1}\rangle\!\rangle\!\langle\!\langle{#2}|}}

\DeclareMathOperator{\Tr}{Tr}

\renewcommand\L{\mathcal{L}}

\newcommand{\M}{\mathcal{M}}
\newcommand{\HS}{\mathcal{H}}

\newcommand{\id}{\mathbbm{1}}

\newcommand{\pA}{{\tilde{A}}}
\newcommand{\pB}{{\tilde{B}}}
\newcommand{\pC}{{\tilde{C}}}
\newcommand{\pD}{{\tilde{D}}}
\newcommand{\pAI}{{\tilde{A}_I}}
\newcommand{\pBI}{{\tilde{B}_I}}

\newcommand{\pDI}{{\tilde{D}_I}}
\newcommand{\pAO}{{\tilde{A}_O}}
\newcommand{\pBO}{{\tilde{B}_O}}

\newcommand{\pDO}{{\tilde{D}_O}}
\newcommand{\pP}{{\tilde{P}}}

\newcommand{\pM}{{\mathsf{M}}}
\newcommand{\pE}{{\mathsf{E}}}
\newcommand{\rA}{{A'}}
\newcommand{\rB}{{B'}}
\newcommand{\rC}{{C'}}
\newcommand{\rD}{{D'}}
\newcommand{\rAI}{{A_I'}}
\newcommand{\rBI}{{B_I'}}

\newcommand{\rAO}{{A_O'}}
\newcommand{\rBO}{{B_O'}}

\newcommand{\rP}{{P'}}
\newcommand{\rF}{{F'}}

\newcommand{\jA}{A^{\textup{j}}}

\newcommand{\jC}{C^{\textup{j}}}

\newcommand{\fA}{A^{\textup{f}}}

\newcommand{\fC}{C^{\textup{f}}}

\newcommand{\jfA}{A^{\textup{jf}}}
\newcommand{\jfB}{B^{\textup{jf}}}
\newcommand{\jfC}{C^{\textup{jf}}}

\newcommand{\dimd}{\textit{\textsf{d}}}

\newcommand{\changed}[1]{{\color{red} #1}}

\newcommand{\notext}[1]{}

\makeatletter 
    
\renewcommand\onecolumngrid{
\do@columngrid{one}{\@ne}%
\def\set@footnotewidth{\onecolumngrid}
\def\footnoterule{\kern-6pt\hrule width 1.5in\kern6pt}%
}

\renewcommand\twocolumngrid{
        \def\footnoterule{
        \dimen@\skip\footins\divide\dimen@\thr@@
        \kern-\dimen@\hrule width.5in\kern\dimen@}
        \do@columngrid{mlt}{\tw@}
}%

\makeatother    

\begin{document}

\title{Network-Device-Independent Certification of Causal Nonseparability}

\author{Hippolyte Dourdent}
\affiliation{ICFO-Institut de Ciencies Fotoniques, The Barcelona Institute of Science and Technology, 08860 Castelldefels, Barcelona, Spain}

\author{Alastair A.\ Abbott}
\affiliation{Univ.\ Grenoble Alpes, Inria, 38000 Grenoble, France}

\author{Ivan Šupić}
\affiliation{LIP6, Sorbonne Université, CNRS, 4 Place Jussieu, 75005 Paris, France}

\author{Cyril Branciard}
\affiliation{Univ.\ Grenoble Alpes, CNRS, Grenoble INP, Institut N\'eel, 38000 Grenoble, France}


\begin{abstract}
Causal nonseparability is the property underlying quantum processes incompatible with a definite causal order. So far it has remained a central open question as to whether any process with a clear physical realisation can violate a causal inequality, so that its causal nonseparability can be certified in a device-independent way, as originally conceived. 
Here we show how a device-independent approach, based solely on the observed correlations, can certify the causal nonseparability of all the processes that can induce a causally nonseparable distributed measurement in a scenario with trusted quantum input states, as defined in [Dourdent \emph{et al.}, \href{https://doi.org/10.1103/PhysRevLett.129.090402}{Phys.\ Rev.\ Lett.\ \textbf{129}, 090402 (2022)}]. 
This notably includes the celebrated quantum switch. This device-independent certification is achieved by introducing a network of untrusted operations, allowing one to self-test the quantum inputs on which the effective distributed measurement induced by the process is performed. 
\end{abstract}

\maketitle


\section{Introduction}

For both quantum and classical processes, it is usually assumed that if two parties interact only once with a physical medium, then only one-way influences are possible. 
In such a process, the correlations the parties can establish are subsequently restricted by this causal structure. Remarkably, this turns out to be an unnecessary assumption.
Indeed, by studying processes where quantum theory is taken to hold locally but no global causal structure is assumed, one finds that
some processes allow parties to establish correlations incompatible with a definite causal order~\cite{oreshkov12}.

Such processes can be studied formally in the process matrix framework~\cite{oreshkov12}, and their lack of a well-defined underlying causal structure formalised through the property of causal nonseparability~\cite{oreshkov12,wechs19}.
Causally nonseparable processes are of foundational interest, for example in the study of quantum gravity where the space-time structure itself may be quantum and hence indefinite~\cite{hardy07}, but also as they have been shown to lead to
advantages in a number of different tasks~\cite{chiribella13,chiribella12,araujo14,guerin16,ebler18,zhao20}.
This role as a potential new quantum resource, as well as ongoing uncertainty about the physical realisability of certain causally nonseparable processes~\cite{araujo17}, means it is important to be able to certify the causal nonseparability of a process.

The strongest version of such certification would be obtained through a \textit{device-independent} (DI) approach, utilising only the observed statistics and based on the violation of ``causal inequalities'' by noncausal correlations~\cite{oreshkov12,branciard16}. 
However, not all causally nonseparable process matrices are noncausal in this strong sense~\cite{araujo15,oreshkov16,feix16}.
Indeed, a large class of quantum-realisable processes cannot violate causal inequalities~\cite{wechs21,purves21}, and it is still unclear whether any process with a clear physical realisation is able to. This includes in particular the canonical ``quantum switch''~\cite{chiribella13}, the resource behind most known advantages arising from causal indefiniteness~\cite{chiribella13,chiribella12,araujo14,guerin16,ebler18,zhao20}.

In a previous work~\cite{dourdent21}, we showed (with another colleague) that the quantum switch, as well as a large class of causally nonseparable processes in both the bipartite scenario and a restricted tripartite one (and even all bipartite scenarios under a reasonable additional assumption), can display a new form of noncausality in a semi-DI with quantum inputs (SDI-QI) scenario. 
In the present work, we show how the gap between this SDI-QI certification and a fully DI certification can be bridged by self-testing the quantum inputs in a network scenario, where each party that receives a quantum input in the SDI-QI scenario shares an additional uncharacterised bipartite state with an extra untrusted party. 
We call this certification \textit{network-device-independent (NDI)} to differentiate it from the standard DI scenario based on causal inequalities, as we assume a specific network structure, i.e., the connectivity between certain independent, uncharacterised quantum devices.

Our approach is motivated by the ability to exploit self-testing to transform a universal certification of entanglement in a SDI-QI scenario into an NDI certification~\cite{bowles18,bowles18a}. 
The apparent analogy between entanglement and causal nonseparability, however, was already seen to have severe limitations in the SDI-QI scenario~\cite{dourdent21}.
Indeed, while all entangled states can be certified in a ``measurement-device-independent'' way (i.e., in an SDI-QI scenario), no analogous certification is known for causally non-separable processes. 
Such limitations arise from the fact that the certification of causal nonseparability involves more complex and structured objects (causally (non)separable processes and quantum instruments) than for entanglement ((non)separable states and measurements), making it unclear whether NDI certification could be applied to this latter resource. Our approach addresses and dispels these doubts, showing that any process whose causal nonseparability can be certified in an SDI-QI manner also admits an NDI certification.


\section{Causal (non)separability in the SDI-QI scenario}

The approach we present is a general method for transforming SDI-QI certification of causal nonseparability into NDI certification, applicable to a range of scenarios.
For concreteness and clarity, however, we focus our initial presentation here on the bipartite scenario, before making some comments about its applicability to other scenarios and processes later.

Two parties, Alice and Bob, control separate quantum labs with input and output Hilbert spaces $\HS^{A_I}$ and $\HS^{A_O}$ for Alice, and $\HS^{B_I}$ and $\HS^{B_O}$ for Bob.
They may also receive some auxiliary quantum states in Hilbert spaces $\HS^{\rA},\HS^{\rB}$.%
\footnote{All Hilbert spaces we consider are taken to be finite-dimensional.
Notations: $\L(\HS^X)$ denotes the space of linear operators on $\HS^X$; superscripts are used to indicate on what spaces operators act; we shall write concisely $\HS^{XY} = \HS^X\otimes\HS^Y$, $\HS^A = \HS^{A_IA_O}$, etc.\\
Note that in~\cite{dourdent21}, the auxiliary spaces in which the quantum inputs live were denoted with tildes (e.g., $\HS^\pA$); here we switch to primes: the tildes will be used for the untrusted physical systems in the self-testing procedure below. Also, in~\cite{dourdent21} Alice and Bob's quantum input states were labeled by subscripts $x$ and $y$; here we change these to $z$ and $w$, as $x$ and $y$ will be introduced later as classical inputs for the self-testing procedure.}
Let us consider the SDI-QI scenario introduced in~\cite{dourdent21} where Alice and Bob are provided with quantum input states $\rho_z^\rA \in\L(\HS^\rA)$ and $\rho_w^\rB \in\L(\HS^\rB)$, respectively, indexed by the labels $z$ and $w$.
They each perform some fixed quantum operations described as quantum instruments~\cite{davies70}, i.e., sets of completely positive (CP) maps $\M_a: \L(\HS^{\rA A_I})\to\L(\HS^{A_O})$ and $\M_b: \L(\HS^{\rB B_I})\to\L(\HS^{B_O})$, whose indices $a,b$ refer to some (classical) outcomes for Alice and Bob, and whose sums $\sum_a \M_a$ and $\sum_b \M_b$ are trace-preserving (TP). Using the Choi isomorphism~\cite{choi75} (see the Supplemental Material (SM) of~\cite{dourdent21}, Sec.~A), the CP maps $\M_a$, $\M_b$ can be represented as positive semidefinite matrices $M_a^{\rA A}$ and $M_b^{\rB B}$.

\begin{figure}[t]
	\begin{center}
	\includegraphics[width=0.7\columnwidth]{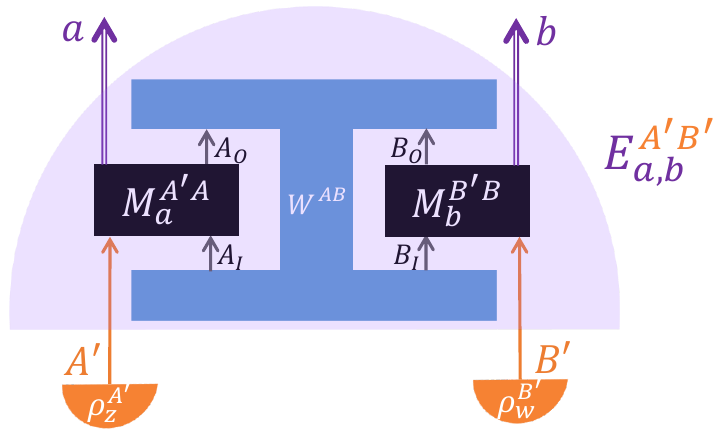}
	\end{center}
	\caption{The SDI-QI scenario: A process matrix $W^{AB}$ connects two parties who receive trusted (orange) quantum inputs $\rho_z^\rA$ and $\rho_w^\rB$, resp.
	They each perform an untrusted (black) joint operation ($(M_a^{\rA A})_a$ and $(M_b^{\rB B})_b$, resp.), producing the classical outcomes $a$ and $b$.
	The purple semicircle represents the D-POVM $(E_{a,b}^{\rA\rB})_{a,b}$ induced by these instruments and the process matrix.} 
	\label{fig:scenarii0}
\end{figure}

Within the process matrix framework, the correlations established by Alice and Bob in such a scenario with quantum inputs are given by the probabilities 
\begin{align}
& P(a,b|\rho_z^\rA,\rho_w^\rB) \notag\\
&\!=\! \big(M_a^{\rA A}\otimes M_b^{\rB B}\big) * \big(\rho_z^\rA\!\otimes\rho_w^\rB\!\otimes W^{AB}\big) \notag \\
& =  E_{a,b}^{\rA\rB} \!*\! \big(\rho_z^\rA\!\!\otimes\rho_w^\rB\big)
 = \Tr \left[ \big(E_{a,b}^{\rA\rB}\big)^T \big(\rho_z^\rA\!\!\otimes\rho_w^\rB\big) \right]
\label{eq:Pab_rhoxrhoy}
\end{align}
with
\begin{align}
\label{eq:D-POVM_def}
E_{a,b}^{\rA\rB} = \big(M_a^{\rA A}\otimes M_b^{\rB B}\big)*W^{AB}.
\end{align}
Here, $W^{AB} \in \L(\HS^{AB})$ is the so-called ``process matrix'', a positive semidefinite matrix satisfying nontrivial linear constraints in order to always generate valid probabilities~\cite{oreshkov12} (see e.g.~\cite{dourdent21}, SM, Sec.~B); and $*$ is the ``link product''~\cite{chiribella08,chiribella09}, a convenient tool for calculations defined for any matrices $M^{XY} \in \L(\HS^{XY})$, $N^{YZ} \in \L(\HS^{YZ})$ as $M^{XY}*N^{YZ} = \Tr_Y[(M^{XY}\otimes\id^Z)^{T_Y}(\id^X\otimes N^{YZ})] \in \L(\HS^{XZ})$ (where $T_Y$ is the partial transpose over $\HS^Y$; see also~\cite{dourdent21}, SM, Sec.~A).\footnote{A full trace $\Tr[(M^Y)^T N^Y]$ and a tensor product $M^X \otimes N^Z$ can both be written as a link product. Moreover, the link product is commutative and associative.} 
The family $\mathbb{E}^{\rA\rB} \coloneqq (E_{a,b}^{\rA\rB})_{a,b}$ defines an effective distributed measurement~\cite{supic17,hoban18} on the quantum inputs, termed a ``distributed positive-operator-valued measure'' (D-POVM); see Fig.~\ref{fig:scenarii0}. 
Note that by choosing $\{\rho_z^\rA\}_z$ and $\{\rho_w^\rB\}_w$ to be tomographically complete sets of states, one can completely reconstruct $\mathbb{E}^{\rA\rB}$ from the statistics obtained through Eq.~\eqref{eq:Pab_rhoxrhoy}.

The process matrix formalism makes no \emph{a priori} assumption of a global causal structure relating Alice and Bob.
In fact, assuming such a structure imposes further constraints, due to the inability for a party to signal to the causal past.
Process matrices compatible, for example, with Alice acting before Bob (denoted $A\prec B$) are of the form $W^{A\prec B} = W^{A\prec B_I} \otimes \id^{B_O}$, and similarly $W^{B\prec A} = W^{B\prec A_I} \otimes \id^{A_O}$ for Bob before Alice ($B\prec A$), with $W^{A\prec B_I} \in \L(\HS^{AB_I})$ and $W^{B\prec A_I} \in \L(\HS^{A_IB})$ being themselves valid process matrices~\cite{oreshkov12}.
Process matrices that can be written as a convex mixture of matrices compatible with $A\prec B$ and $B\prec A$, i.e., of the form
\begin{align}
W^{AB} & = q\,W^{A\prec B_I} \otimes \id^{B_O} + (1{-}q)\,W^{B\prec A_I} \otimes \id^{A_O},
\label{eq:csep2}
\end{align}
with $q\in[0,1]$, are said to be ``causally separable''. 
They can be interpreted as being compatible with a definite (although perhaps probabilistic) causal order. 
Remarkably, there exist ``causally nonseparable'' process matrices that cannot be decomposed as in Eq.~\eqref{eq:csep2}, and are thus incompatible with any definite causal order~\cite{oreshkov12}.

Causal nonseparability can always be certified in a device-dependent manner using a ``causal witness''~\cite{araujo15,branciard16a}, while only some processes can be certified in a DI way through the violation of a causal inequality~\cite{oreshkov12,araujo15,oreshkov16,feix16}. 
Here, we start by considering a relaxation of this DI scenario, by providing the parties with quantum inputs instead of classical ones. 
In~\cite{dourdent21}, we showed that causally separable processes necessarily generate D-POVMs that are themselves causally separable in the sense that they can be written
\begin{align}
\mathbb{E}^{\rA\rB} = q\, \mathbb{E}^{\rA\prec\rB} + (1{-}q)\, \mathbb{E}^{\rB\prec\rA} \label{eq:csep_DPOVM}
\end{align}
with $q\in[0,1]$, where  $\mathbb{E}^{\rA\prec\rB} = (E_{a,b}^{\rA\prec\rB})_{a,b}$ and $\mathbb{E}^{\rB\prec\rA} = (E_{a,b}^{\rB\prec\rA})_{a,b}$ are D-POVMs satisfying
\begin{align}
    \sum_b E_{a,b}^{\rA\prec\rB} = E_a^\rA \otimes \id^\rB\hspace{3mm} \forall a, \hspace{3mm} \sum_a E_a^\rA=\id^\rA,\label{eq:dpovmab}\\
    \sum_a E_{a,b}^{\rB\prec\rA} = E_b^\rB \otimes \id^\rA\hspace{3mm} \forall b,\hspace{3mm} \sum_b E_b^\rB=\id^\rB,\label{eq:dpovmba}
\end{align} 
and are respectively compatible with the causal order where Alice receives her quantum input and acts before Bob ($\rA\prec\rB$) and \emph{vice versa} ($\rB\prec\rA$).
Crucially, some causally nonseparable processes which cannot violate any causal inequality in the DI setting (such as the ``causal'' process of Ref.~\cite{feix16}) can still generate causally nonseparable D-POVMs. 
This causal nonseparability---and hence that of the process matrix inducing it through Eq.~\eqref{eq:D-POVM_def}---can then be certified by reconstructing the D-POVM (as mentioned above) and verifying that it cannot be decomposed as in Eq.~\eqref{eq:csep_DPOVM} or, more directly, using these quantum inputs to violate a witness inequality of the form $\mathcal{J}\ge 0$, where $\mathcal{J}$ denotes some linear combination of the probabilities $P(a,b|\rho_z^\rA,\rho_w^\rB)$~\cite{dourdent21}.
Note that no characterisation is currently known of the process matrices for which such a SDI-QI certification of causal nonseparability is possible. 
However, unlike the case for entanglement---where a simple recipe provides measurements for all parties that give rise to a SDI-QI (or ``measurement-DI'') certification of any entangled state~\cite{buscemi12,branciard13}---it seems that no such simple recipe for inducing causally nonseparable D-POVMs exists, and it was conjectured that some process matrices do not admit a SDI-QI certification~\cite{dourdent21}.


\section{NSDI-QI and NDI causal scenarios}

\begin{figure}[t]
	\begin{center}
	\includegraphics[width=0.8\columnwidth]{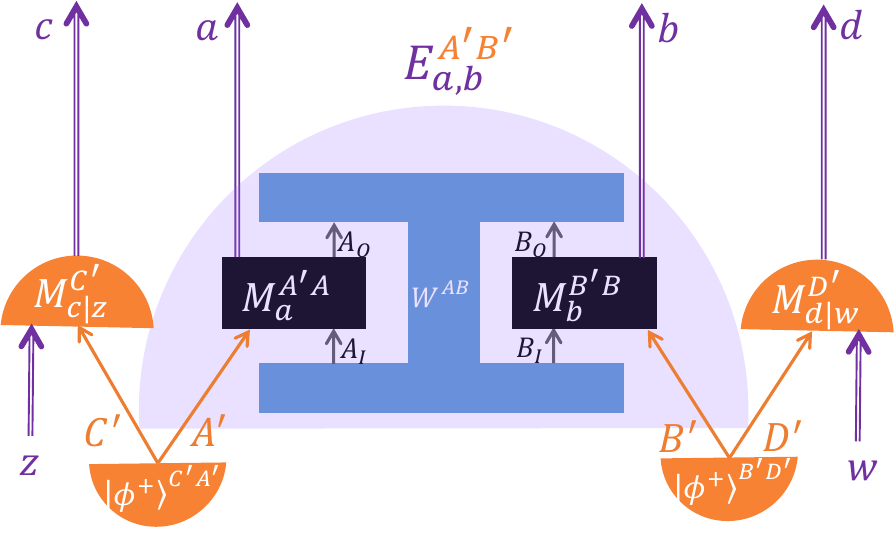}\\
	\vspace{5mm}
	\includegraphics[width=0.8\columnwidth]{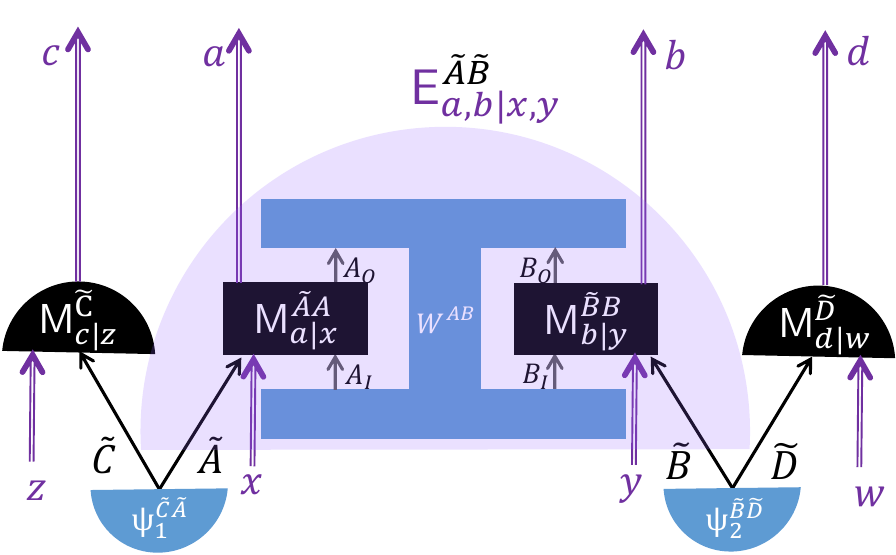}
	\end{center}
	\caption{From the NSDI-QI to the NDI scenario: In a NSDI-QI scenario (top), the quantum inputs of the SDI-QI scenario (Fig.~\ref{fig:scenarii0}) are prepared remotely by two other separated parties---Charlie and Daisy---performing suitable operations on their part of a maximally entangled state shared with Alice and Bob, so that $\rho_z^\rA\propto M_{0|z}^\rC*\ketbra{\phi^+}{\phi^+}^{\rC\rA}$ and $\rho_w^\rB\propto M_{0|w}^\rD*\ketbra{\phi^+}{\phi^+}^{\rB\rD}$ resp.\ (orange). 
	These quantum inputs can be self-tested (bottom): Alice, Bob, Charlie and Daisy are given classical inputs $x,y,z,w$, and perform untrusted operations that produce outcomes $a,b,c,d$, resp. 
	Verifying that the correlations $P_y(c,a|z,x)$ and $P_x(b,d|y,w)$ both maximally violate a Bell inequality allows one to self-test the preparation of the quantum inputs; from this, one can certify the causal nonseparability of the D-POVM $(\mathsf{E}_{a,b|x,y}^{\pA\pB})_{a,b}$ (purple semicircle) for $x=y=\star$, and thus the causal nonseparability of the process matrix $W^{AB}$, in a  NDI scenario. 
 }
	\label{fig:scenarii}
\end{figure}

A bipartite SDI-QI scenario can be transformed into a 4-partite network scenario (NSDI-QI) where two additional separated parties---Charlie and Daisy---remotely prepare the quantum inputs. 
They control separate labs with input Hilbert spaces $\HS^\rC$ and $\HS^\rD$, and share maximally entangled pairs of $\dimd$-dimensional states,%
\footnote{Here $\dimd$ is the dimension of the quantum input spaces, which for simplicity we assume to be the same for both parties (although this can be easily relaxed without consequence).}
$\ket{\phi^+}^{\rC\rA}$ and $\ket{\phi^+}^{\rB\rD}$ (with $\ket{\phi^+} = \frac{1}{\sqrt{\dimd}}\sum_{j=0}^{\dimd-1}\ket{j,j}$) with Alice and Bob, respectively; see Fig.~\ref{fig:scenarii} (top).
Charlie and Daisy can then remotely prepare the quantum inputs $\rho_z^\rA$ and $\rho_w^\rB$ by applying measurements with Choi representations $(M_{c|z}^\rC)_{c} = (\rho_z^\rC, \id^\rC-\rho_z^\rC)$ and $(M_{d|w}^\rD)_{d} = (\rho_w^\rD, \id^\rD-\rho_w^\rD)$ on their shares of entangled qudit pairs.
When they observe the outcomes $c=0$ and $d=0$ corresponding to the POVM elements $\rho_{z}^\rC$ and $\rho_{w}^\rD$, respectively, then Alice and Bob receive precisely the desired quantum inputs since $\rho_{z}^\rA\propto M_{0|z}^\rC*\ketbra{\phi^+}{\phi^+}^{\rC\rA}$ and $\rho_{w}^\rB\propto M_{0|w}^\rD*\ketbra{\phi^+}{\phi^+}^{\rB\rD}$.
Thereby, the observed correlations in this network scenario are
\begin{align}
    P(c,a,b,d|z,w) =&   \left(M_{c|z}^\rC \otimes E_{a,b}^{\rA\rB}\otimes M_{d|w}^\rD\right) \notag\\
    &* \left(\ketbra{\phi^+}{\phi^+}^{\rC\rA}\otimes\ketbra{\phi^+}{\phi^+}^{\rB\rD}\right) \label{eq:P_cabd_zw}
\end{align}
and satisfy $P(0,a,b,0|z,w) = P(0|z)P(0|w)P(a,b|\rho_z^\rA,\rho_w^\rB)$. The causal nonseparability of $\mathbb{E}^{\rA\rB}$ can be certified by the violation of a witness inequality as in the SDI-QI scenario. 

This NSDI-QI scenario so far assumes the sharing of maximally entangled states and trusts Charlie's and Daisy's quantum measurements in the ``reference scenario'' described above. 
However, these assumptions can be relaxed by exploiting self-testing techniques. 
Self-testing exploits the fact that quantum correlations which maximally violate a Bell inequality can sometimes be used to uniquely determine---up to local transformations---which state and measurements produced them~\cite{mayers04,supic20a}.
To self-test the maximally entangled states and Charlie and Daisy's measurements, Alice, Bob, Charlie and Daisy are all given (new) classical inputs $x,y,z,w$, resp., and perform untrusted operations $(\mathsf{M}_{a|x}^{\pA A})_{a}, (\mathsf{M}_{b|y}^{\pB B})_{b}, (\mathsf{M}_{c|z}^\pC)_{c}, (\mathsf{M}_{d|w}^\pD)_{d}$, producing outcomes $a,b,c,d$, resp.\ on the process matrix $W^{AB}$ and the untrusted auxiliary states $\uppsi_1^{\pC\pA}$ and $\uppsi_2^{\pB\pD}$ to be self-tested, which are now in the (untrusted) ``physical'' Hilbert spaces $\HS^{\pA},\HS^{\pB},\HS^{\pC},\HS^{\pD}$ (denoted with tildes)---that, \emph{a priori}, may differ from those in the reference scenario $\HS^{\rA},\HS^{\rB},\HS^{\rC},\HS^{\rD}$ (with primes); see Fig.~\ref{fig:scenarii} (bottom). 
In the spirit of the ``black-box'' DI approach, we assume that the network structure of Fig.~\ref{fig:scenarii} holds, and that the uncharacterised but independent devices nonetheless obey quantum theory.
This network assumption, defined here for quantum theory, implies that the observed correlations can be written as
\begin{align}
P(c,a,b,d|z,x,y,w)  \!=\! & \left(\!\mathsf{M}_{c|z}^\pC \otimes \mathsf{M}_{a|x}^{\pA A}\otimes \mathsf{M}_{b|y}^{\pB B} \otimes \mathsf{M}_{d|w}^\pD\!\right) \notag\\
& *\! \left(\uppsi_1^{\pC\pA}\!\otimes\! W^{AB}\!\otimes\!\uppsi_2^{\pB\pD}\right).
\label{eq:Pabcd_xyzw_NDI}
\end{align}

When applied to the process matrix, Alice and Bob's instruments then generate a D-POVM $\mathbb{E}_{x,y}^{\pA\pB}=(\mathsf{E}_{a,b|x,y}^{\pA\pB})_{a,b}$ for each $x$ and $y$ with elements $\mathsf{E}_{a,b|x,y}^{\pA\pB}=\big(\mathsf{M}_{a|x}^{\pA A}\otimes \mathsf{M}_{b|y}^{\pB B}\big)*W^{AB}$, similarly to Eq.~\eqref{eq:D-POVM_def}. 
We will show that in this scenario, the causal nonseparability of a D-POVM generated from a causally nonseparable process matrix $W^{AB}$ can be certified solely based on the correlations $P(c,a,b,d|z,x,y,w)$ thus achieving an NDI certification of causal nonseparability. 

\section{NDI certification of causal nonseparability}%
As suggested above, the idea to achieve this goal is to combine the SDI-QI certification with self-testing.
For the protocol we present to function, it will be necessary for the quantum inputs in the initial SDI-QI setting to be chosen such that that the reference measurements $(M_{c|z}^\rC)_c=(\rho_z^\rC,\id^\rC - \rho_z^\rC)$ and $(M_{d|w}^\rD)_d=(\rho_w^\rD, \id^\rD - \rho_w^\rD)$ of the NSDI-QI setting are precisely those involved in the self-testing procedure we use.
For simplicity, let us assume that $\dimd=2^m$ for some $m\ge 1$. This can be done without loss of generality by embedding the reference experiment, i.e., the spaces $\HS^{\rC}$ and $\HS^{\rD}$, into larger $2^m$-dimensional spaces if needed (see Appendix~\ref{app:st}). 
We then require not only that $\rho_z^\rC$ and $\rho_w^\rD$ form tomographically complete sets, but more precisely that they are (projectors onto the) eigenstates of orthogonal Pauli measurements (or, beyond the 2-dimensional case, tensor products thereof).
We will assume henceforth that this is indeed the case. 

When Alice and Bob receive generic classical inputs $x,y$ (other than the specific $x,y= \star$ that we shall consider below), we interpret the experiment as a self-testing round, in which the correlations are used to test a Bell inequality: reaching its maximal quantum violation certifies that the untrusted states $\uppsi_1^{\pC\pA},\uppsi_2^{\pB\pD}$ and measurements $(\mathsf{M}_{c|z}^\pC)_{c}, (\mathsf{M}_{d|w}^\pD)_{d}$ have the form of those used in the NSDI-QI setup, potentially up to some local transformations that leave the correlations unchanged (e.g., a change of basis).
When they both receive the special input $x=y=\star$, we instead interpret it as a certification round, in which one follows the same procedure as in a NSDI-QI certification of causal nonseparability using these self-tested quantum inputs.
Crucially, the aforementioned local transformations that formally relate the ``physical'', uncharacterised states and measurements to their idealised ``reference'' counterparts can be explicitly incorporated into our analysis, and shown to preserve causal (non)separability, allowing the causal nonseparability of the physical process to be certified, despite lacking any knowledge of the Hilbert spaces upon which it acts. 

The general protocol can thus be described as follows.
\begin{itemize}
    \item (\textit{Correlation generation}) Alice, Bob, Charlie and Daisy perform local operations and measurements on their respective subsystems to obtain the statistics $P(c, a, b, d|z, x, y, w)$.
    \item (\textit{Self-testing}) One certifies that the auxiliary states each contain a maximally entangled state and that Charlie and Daisy each effectively perform the desired tomographically complete, orthogonal Pauli measurements on their subsystems by verifying that the distributions $P_y(c, a|z, x):=P(c,a|z,x,y)$ and $P_x(b,d|y,w):=P(b, d|x,y, w)$ both maximally violate a Bell inequality $\mathcal{I}\ge 0$ for some $y$ and $x$, resp.%
    \footnote{Note that these distributions are independent of $w$ and $z$, respectively, due to the network structure~\changed{\eqref{eq:Pabcd_xyzw_NDI}}: e.g., Daisy only shares a (nonsignalling) quantum state with Alice, Bob and Charlie. However, contrary to the scenario of entanglement certification~\cite{bowles18,bowles18a}, these may still depend on $y$ and $x$, resp., as the process matrix may allow for signalling.} 
\item (\textit{Causal nonseparability certification}) One verifies that the correlations $P_\star(a,b|z,w):=P(a,b|z, x{=}\star, y{=}\star, w, c{=}0, d{=}0)$ violate a witness inequality $\mathcal{J}\geq 0$, thus certifying that the process matrix is causally nonseparable.
\end{itemize}

Let us now explain these steps in more detail.
We use the fact that, for $\dimd=2^m$, one can self-test a $\dimd$-dimensional maximally entangled state and orthogonal Pauli measurements (or tensor products thereof, for $\dimd>2$) by the maximal quantum violation of a Bell inequality $\mathcal{I}\ge 0$~\cite{bowles18a}.
Further discussion of the inequality is provided in Appendix~\ref{app:st}, but its exact form (which depends on $\dimd$) is not important for the argument here.

The first step of the protocol simply consists in generating the correlations $P(c,a,b,d|z,x,y,w)$, which, developing Eq.~\eqref{eq:Pabcd_xyzw_NDI}, can be written as
\begin{align}
& P(c,a,b,d|z,x,y,w) \notag \\
& \!=\! \left(\!\mathsf{M}_{c|z}^\pC \otimes \mathsf{M}_{a|x}^{\pA A}\otimes \mathsf{M}_{b|y}^{\pB B} \otimes \mathsf{M}_{d|w}^\pD\!\right) \!*\! \left(\uppsi_1^{\pC\pA}\!\otimes\! W^{AB}\!\otimes\!\uppsi_2^{\pB\pD}\right) \notag \\
& = P(c|z)\, P(d|w)\, \mathsf{E}_{a,b|x,y}^{\pA\pB} * \left( \uprho_{c|z}^\pA \otimes \uprho_{d|w}^\pB \right)
\label{eq:Pabcd_xyzw}
\end{align}
with the remotely prepared states satisfying $P(c|z)\, \uprho_{c|z}^\pA = \mathsf{M}_{c|z}^\pC * \uppsi_1^{\pC\pA}$ and $P(d|w)\, \uprho_{d|w}^\pB = \mathsf{M}_{d|w}^\pD * \uppsi_2^{\pB\pD}$, and with the D-POVM elements $\mathsf{E}_{a,b|x,y}^{\pA\pB}=\big(\mathsf{M}_{a|x}^{\pA A}\otimes \mathsf{M}_{b|y}^{\pB B}\big)*W^{AB}$, as given above.

The second step is the self-testing procedure. 
The players verify that the correlations $P_y(c,a|z,x)$ and $P_x(b,d|y,w)$ reach the maximum quantum violation of the Bell inequality $\mathcal{I}\ge 0$.
As we recall in Appendix~\ref{app:st}, it is known that if this is the case, then the physical states $\uppsi_1^{\pC\pA}$ and $\uppsi_2^{\pB\pD}$ and measurements $(\mathsf{M}_{c|z}^\pC)_c$ and $(\mathsf{M}_{d|w}^\pD)_d$ can be related to the ideal reference states and operations of the NSDI-QI scenario in the $\dimd$-dimensional spaces $\HS^\rC,\HS^\rA,\HS^\rB,\HS^\rD$ via local isometries~\cite{bowles18,bowles18a}.
We then show (see Appendix~\ref{app:st}) that these local isometries also relate the remotely prepared inputs%
\footnote{In fact, we show that each $\uprho_{c|z}^\pA$ and $\uprho_{d|w}^\pB$ (for all $c,d$) can be related to a corresponding state in the reference scenario, but this stronger result is not needed here.}
$\uprho_{0|z}^\pA$ and $\uprho_{0|w}^\pB$ to the reference quantum inputs $\rho_{z}^\rA\propto M_{0|z}^\rC*\ketbra{\phi^+}{\phi^+}^{\rC\rA}$ and $\rho_{w}^\rB\propto M_{0|w}^\rD*\ketbra{\phi^+}{\phi^+}^{\rB\rD}$.
For example, for Alice's quantum inputs we prove that
\begin{equation}
		\uprho_{0|z}^\pA = {V^{\mathbf{A}}}^\dagger\left( \rho_z^\rA\otimes\xi_0^{\jfA} + (\rho_z^\rA)^T\otimes\xi_1^{\jfA} \right) V^{\mathbf{A}},
\end{equation}
where $V^{\mathbf{A}}:\HS^\pA\to \HS^{\rA\jfA}$ is an isometry and $\xi_0^{\jfA}$, $\xi_1^{\jfA}$ are (subnormalised) ``junk and flag'' states on a Hilbert space $\HS^{\jfA}$; a similar relation holds for Bob.
This allows us to write Eq.~\eqref{eq:Pabcd_xyzw}, when $c=d=0$, in terms of the reference scenario as
\begin{align}
	\label{eq:Pabcd_xyzw_reference}
	&P(0,a,b,0|z,x,y,w)\notag\\
	& \qquad =P(0|z)\,P(0|w)\,E_{a,b|x,y}^{\rA\rB} * \big( \rho_z^\rA\otimes \rho_w^\rB \big),
\end{align}
where the $E_{a,b|x,y}^{\rA\rB}$ are elements of an effective D-POVM (which $\mathsf{E}_{a,b|x,y}^{\pA\pB}$ can likewise be related to through local isometries; see Eq.~\eqref{eq:rE_from_pE} in Appendix~\ref{app:CNSep_NDI}).
Eq.~\eqref{eq:Pabcd_xyzw_reference} thus relates directly the observed correlations to those obtainable in the NSDI-QI setup.

The last step consists in certifying causal nonseparability. 
The self-testing step ensures that, when Charlie and Daisy get the outputs $c=d=0$, Alice and Bob receive, up to local isometries, the desired quantum inputs $\rho_z^\rA$ and $\rho_w^\rB$ from the SDI-QI scenario.
We thus focus on the correlations observed when $x=y=\star$ which, in the reference experiment, should correspond to the measurements used to obtain a causally non-separable D-POVM.
These correlations, following Eq.~\eqref{eq:Pabcd_xyzw_reference}, are given by
\begin{align}
    P_\star(a,b|z,w) & = P(a,b|z, x{=}\star, y{=}\star, w, c{=}0, d{=}0) \notag \\
    & = E_{a,b|\star,\star}^{\rA\rB} *  \big( \rho_z^\rA\otimes \rho_w^\rB \big), \label{eq:def_Pstar}
\end{align}
which precisely defines, following Eq.~\eqref{eq:Pab_rhoxrhoy}, a distribution $P'(a,b|\rho_z^\rA,\rho_w^\rB)$ in a SDI-QI setting with the reference quantum inputs for the certification of the effective D-POVM $\mathbb{E}_\text{eff}^{\rA\rB}:=(E_{a,b|\star,\star}^{\rA\rB})_{a,b}$. 
In Appendix~\ref{app:CNSep_NDI} we show that if $W^{AB}$ is causally separable then $\mathbb{E}_\text{eff}^{\rA\rB}$ is a causally separable D-POVM.
Hence, if $P'(a,b|\rho_z^\rA,\rho_w^\rB)$ violates the witness inequality, yielding $\mathcal{J} < 0$, this certifies the causally nonseparability of $W^{AB}$.
By following the ideal reference protocol of the NSDI-QI scenario, one can thus obtain such an NDI certification for any process that can be certified in the bipartite SDI-QI scenario, i.e., which can generate a causally nonseparable D-POVM.

The NDI certification of causal nonseparability described above assumes that the self-testing is perfect: the maximal quantum violation of the Bell inequality $\mathcal{I}\ge 0$ is obtained, meaning that noiseless maximally entangled states and perfect measurements are realised (up to local isometries).
Such requirements are experimentally unobtainable. However, the self-testing procedure and subsequent certification of causal nonseparability can be rendered robust to noise in a similar way to that of~\cite{bowles18a}; see Appendix~\ref{app:robustness} for further explanation.


\section{Applications and scope}

The protocol described in the previous section exploits a general approach to transforming a SDI-QI scenario into an NDI one. 
Indeed, it can be readily generalised to provide NDI certification in settings beyond the basic bipartite one we described.
For instance, in~\cite{dourdent21} we showed that all bipartite causally nonseparable process matrices can be certified in a SDI-QI way if we assume an additional natural assumption on the form of Alice's and Bob's instruments: that they factorise as a joint measurement on one subsystem of their quantum input and the system they receive from the process matrix, and a channel taking the rest of their quantum input into the process matrix.
This ``measurement device and channel independent'' (MDCI) structure can naturally be translated into the NDI setting by imposing the same structure on the physical instruments $(\pM_{a|x\text{\sout{$\star$}}}^{\pA A})_a$ and $(\pM_{b|y\text{\sout{$\star$}}}^{\pB B})_b$, corresponding to assuming that Alice and Bob each control two uncharacterised quantum black-box devices that are connected in a specific way. 
This amounts to an additional structural assumption on the network, but the certification remains DI in spirit. 
By using independent auxiliary sources to self-test Alice and Bob's quantum input subsystems, our method here can readily be used to show that all bipartite causally nonseparable processes can also be certified in an NDI way (modulo these assumptions); see Appendix~\ref{app:MDCI} for details on the soundness of the NDI certification in this MDCI scenario.

A process of particular interest is the quantum switch~\cite{chiribella13}, which 
can be described in either a ``($2$+$F$)-partite'' or ``($P$+$2$+$F$)-partite scenario'' in which, in addition to Alice and Bob, a third party Fiona with no output Hilbert space performs a measurement in their causal future, and (in the ($P$+$2$+$F$)-partite scenario), a fourth party Phil with no input Hilbert space acts in their causal past.
In the quantum switch, Phil prepares a target system and a control qubit, the latter of which coherently controls the order that Alice and Bob then act on the target system, before both the target and control are received by Fiona.
(In the ($2$+$F$)-partite scenario, Phil's operations are fixed and absorbed into the process matrix.)
The quantum switch is the canonical example of a physically realisable causally nonseparable process~\cite{wechs21}, and the resource behind most known advantages arising from causal indefiniteness~\cite{chiribella13,chiribella12,araujo14,guerin16,ebler18,zhao20}. 
Certifying its causal nonseparability is thus a key problem, and several experiments have done this in a device-dependent way using causal witnesses~\cite{rubino17,goswami18}.
In~\cite{dourdent21}, the SDI-QI scenario was extended to the ($2$+$F$)-partite scenario to certify the causal nonseparability of the quantum switch, and in Appendix~\ref{app:sdiqiP2F} we further extend it to the ($P$+$2$+$F$)-partite scenario.
One can then transform SDI-QI certifications in these scenarios into NDI ones by also having Fiona and, in the later scenario, Phil, share maximally entangled states with additional parties.
Generalising naturally our bipartite protocol, one can independently self-test the quantum inputs of each party and analogously show that statistics violating a witness inequality in the certification rounds imply that the physical process must be causally nonseparable; see Appendix~\ref{app:sdiqiP2F} for further discussion.
For the quantum switch, in the SDI-QI setting it is in fact not necessary to provide all parties with quantum inputs: some parties may do with classical inputs, or no input at all. For example, in~\cite{dourdent21} the case where only Alice and Bob have quantum inputs was considered; in Appendix~\ref{app:sdiqiP2F} we show that one can also certify the causal nonseparability of the quantum switch in a  SDI-QI manner while providing only Phil with quantum inputs---a scenario that can then be turned into an NDI one by introducing just one additional party.
We thus see that, despite its inability to violate any causal inequality~\cite{araujo15,oreshkov16}, the quantum switch can be certified in a particular type of DI scenario; see also the discussion below.
As in the bipartite scenario, it remains an open question which causally nonseparable processes can be certified in an SDI-QI---and thus NDI---manner in the ($2$+$F$)- and ($P$+$2$+$F$)-partite scenarios.

Beyond the bipartite and restricted three- and four-partite scenarios discussed above, the study of SDI-QI certification of causal nonseparability remains an open question.
Indeed, the notion of causal separability of D-POVMs has thus far only been defined in these scenarios.
Nonetheless, the general method we present for converting a SDI-QI scenario into an NDI one can be applied generally, introducing additional parties remotely preparing the quantum inputs and independently self-testing these quantum inputs in the way we described.
A SDI-QI protocol leading to a D-POVM unobtainable from any causally separable process constitutes a SDI-QI certification of causal nonseparability,\footnote{Here and in Ref.~\cite{dourdent21} the notion of a causally separable D-POVM was defined directly. However, Ref.~\cite{dourdent21} showed that, at least in the bipartite and ($2$+$F$)-partite scenarios, the definition precisely coincides with the set of D-POVMs obtainable from causally separable process matrices.} which could then be transformed in this way into an NDI certification of causal nonseparability by generalising the proof of Appendix~\ref{app:CNSep_NDI} that the self-testing procedure preserves causal separability.
An understanding of which causally nonseparable process can be certified in this way in general multipartite scenarios is currently lacking, but this nevertheless highlights the generality of our recipe for converting SDI-QI certification to NDI certification.


\section{Discussion}

As we emphasised, our certification of causal nonseparability is \textit{device-independent}, in the sense that it does not assume any knowledge about the underlying process or operations involved.
Every element involved in the protocol is treated as a (quantum) black box with classical inputs and outputs, and all conclusions are drawn from the observed correlations. 
However, we called it \textit{network-device-independent} to differentiate it from the standard approach to DI certification, which requires the violation of a causal inequality~\cite{oreshkov12}. 
Indeed, our approach differs from the standard one as it assumes a network structure involving two separated untrusted parties, each independently performing operations on a system that is independently shared with a party involved in the process being certified. 
In the black-box picture this corresponds to an assumption on the connectivity of the boxes allowing systems to be shared, and thereby correlations to be generated.
Moreover, unlike the standard approach which certifies causal indefiniteness in a theory-independent manner, our method certifies the theory-dependent notion of causal nonseparability. 
Indeed, it is based on the violation of a witness inequality by a causally nonseparable process matrix and on the self-testing of quantum input states through the maximal quantum violation of Bell inequalities. 
It thus assumes the validity of quantum theory and of its extension through the process matrix formalism.  
This, and the simple \textit{network} structural assumption, is what allows one to certify---without trusting any device---the causal nonseparability of process matrices which cannot violate a causal inequality, such as the quantum switch~\cite{araujo15,oreshkov16}.

Finally, we note that some recent works have also proposed another approach to certify the indefinite causal order in the quantum switch in a device-independent way~\cite{gogioso23,lugt22,lugt23}.
In the scenario presented  in~\cite{lugt22} (also considered in~\cite{gogioso23}), an additional space-like separated party is introduced along with Alice, Bob and Fiona, and an inequality is derived based on the assumptions of ``Definite causal order, Relativistic causality, and Free interventions'' (DRF).
It is then shown that the correlations generated from a quantum switch in which the control system is entangled with the newly introduced party can violate this ``DRF inequality''.
Notably, our approach allows us to consider a scenario very close to that of~\cite{lugt22} by transforming a SDI-QI test of a ($P$+$2$+$F$)-partite quantum switch, with only Phil (in the causal past) being given a quantum input, into an NDI test (as briefly considered above, and further discussed in Appendix~\ref{app:sdiqiP2F}). 
It is then interesting to compare the assumptions and conclusions one may draw from these two approaches in which the causal indefiniteness of the quantum switch can be certified in a DI manner (see Appendix~\ref{app:DRF} for further discussion). 
A first important difference is that the approach of~\cite{lugt22}, and the inequality they derive, is not only device-independent, but also theory-independent---like standard causal inequalities, but contrary to our approach. 
Therefore, while~\cite{lugt22} explicitly introduced some form of no-signalling assumption at the ontological level (``Relativistic causality'') between the new party and those involved in the switch, our approach, being theory-dependent, does not rely on such an assumption. 
Instead, we make the (quantum) network structure assumption, Eq.~\eqref{eq:Pabcd_xyzw_NDI}, which implies immediately the  operational notion of no-signalling. 
With a violation of a ``DRF inequality'', the conclusion that there is indefinite causal order can hence be escaped by rejecting this relativistic causality assumption; in our quantum theory-dependent case, one would need to reject our network structure assumption to salvage definite causal order (or, in both cases, one could reject the ``Free interventions assumption'', which we indeed also made implicitly). 
It will be insightful to further investigate the connections between the two approaches, possibly in different scenarios. Recall in particular that our approach provides a systematic way to turn a SDI-QI certification of causal nonseparability into an NDI one; it would be interesting to see if the approach of~\cite{lugt22} can similarly be made more systematic.


\paragraph{Acknowledgements}
We thank Jean-Daniel Bancal, Nicolas Brunner and Tein van der Lugt for enlightening discussions, and acknowledge financial support from the EU NextGen Funds, the Government of Spain (FIS2020-TRANQI and Severo Ochoa CEX2019-000910-S), Fundació Cellex, Fundació Mir-Puig, Generalitat de Catalunya (CERCA program), the PEPR integrated project EPiQ ANR-22-PETQ-0007 part of Plan France 2030, the ERC Starting grant QUSCO, and the Agence Nationale de la Recherche (project ANR-22-CE47-0012).

For the purpose of open access, the authors have applied a CC-BY public copyright licence to any Author Accepted Manuscript (AAM) version arising from this submission.


\bibliographystyle{quantum}
\bibliography{biblio.bib}

\begin{thebibliography}{10}

\bibitem{oreshkov12}
Ognyan Oreshkov, Fabio Costa, and {\v{C}}aslav Brukner.
\newblock ``Quantum correlations with no causal order''.
\newblock \href{https://doi.org/10.1038/ncomms2076}{Nat. Commun. {\bf 3}, 1092}~(2012).
\newblock  \href{http://arxiv.org/abs/1105.4464}{arXiv:1105.4464}.

\bibitem{wechs19}
Julian {Wechs}, Alastair~A. {Abbott}, and Cyril {Branciard}.
\newblock ``{On the definition and characterisation of multipartite causal (non)separability}''.
\newblock \href{https://doi.org/10.1088/1367-2630/aaf352}{New J. Phys. {\bf 21}, 013027}~(2019).
\newblock  \href{http://arxiv.org/abs/1807.10557}{arXiv:1807.10557}.

\bibitem{hardy07}
Lucien Hardy.
\newblock ``Towards quantum gravity: a framework for probabilistic theories with non-fixed causal structure''.
\newblock \href{https://doi.org/10.1088/1751-8113/40/12/S12}{J. Phys. A: Math. Theor. {\bf 40}, 3081}~(2007).
\newblock  \href{http://arxiv.org/abs/gr-qc/0608043}{arXiv:gr-qc/0608043}.

\bibitem{chiribella13}
Giulio Chiribella, Giacomo~Mauro D'Ariano, Paulo Perinotti, and Beno{\^\i}t Valiron.
\newblock ``Quantum computations without definite causal structure''.
\newblock \href{https://doi.org/10.1103/PhysRevA.88.022318}{Phys. Rev. A {\bf 88}, 022318}~(2013).
\newblock  \href{http://arxiv.org/abs/0912.0195}{arXiv:0912.0195}.

\bibitem{chiribella12}
Giulio Chiribella.
\newblock ``Perfect discrimination of no-signalling channels via quantum superposition of causal structures''.
\newblock \href{https://doi.org/10.1103/PhysRevA.86.040301}{Phys. Rev. A {\bf 86}, 040301}~(2012).
\newblock  \href{http://arxiv.org/abs/1109.5154}{arXiv:1109.5154}.

\bibitem{araujo14}
Mateus {Ara{\'u}jo}, Fabio {Costa}, and {\v C}aslav {Brukner}.
\newblock ``{Computational Advantage from Quantum-Controlled Ordering of Gates}''.
\newblock \href{https://doi.org/10.1103/PhysRevLett.113.250402}{Phys. Rev. Lett. {\bf 113}, 250402}~(2014).
\newblock  \href{http://arxiv.org/abs/1401.8127}{arXiv:1401.8127}.

\bibitem{guerin16}
Philippe~Allard Gu{\'e}rin, Adrien Feix, Mateus Ara{\'u}jo, and {\v C}aslav Brukner.
\newblock ``Exponential communication complexity advantage from quantum superposition of the direction of communication''.
\newblock \href{https://doi.org/10.1103/PhysRevLett.117.100502}{Phys. Rev. Lett. {\bf 117}, 100502}~(2016).
\newblock  \href{http://arxiv.org/abs/1605.07372}{arXiv:1605.07372}.

\bibitem{ebler18}
Daniel Ebler, Sina Salek, and Giulio Chiribella.
\newblock ``Enhanced communication with the assistance of indefinite causal order''.
\newblock \href{https://doi.org/10.1103/PhysRevLett.120.120502}{Phys. Rev. Lett. {\bf 120}, 120502}~(2018).
\newblock  \href{http://arxiv.org/abs/1711.10165}{arXiv:1711.10165}.

\bibitem{zhao20}
Xiaobin Zhao, Yuxiang Yang, and Giulio Chiribella.
\newblock ``Quantum metrology with indefinite causal order''.
\newblock \href{https://doi.org/10.1103/PhysRevLett.124.190503}{Phys. Rev. Lett. {\bf 124}, 190503}~(2020).
\newblock  \href{http://arxiv.org/abs/1912.02449}{arXiv:1912.02449}.

\bibitem{araujo17}
Mateus Ara{\'{u}}jo, Adrien Feix, Miguel Navascu{\'{e}}s, and {\v{C}}aslav Brukner.
\newblock ``A purification postulate for quantum mechanics with indefinite causal order''.
\newblock \href{https://doi.org/10.22331/q-2017-04-26-10}{{Quantum} {\bf 1}, 10}~(2017).
\newblock  \href{http://arxiv.org/abs/1611.08535}{arXiv:1611.08535}.

\bibitem{branciard16}
Cyril Branciard, Mateus Ara{\'{u}}jo, Adrien Feix, Fabio Costa, and {\v{C}}aslav Brukner.
\newblock ``The simplest causal inequalities and their violation''.
\newblock \href{https://doi.org/10.1088/1367-2630/18/1/013008}{New J. Phys. {\bf 18}, 013008}~(2016).
\newblock  \href{http://arxiv.org/abs/1508.01704}{arXiv:1508.01704}.

\bibitem{araujo15}
Mateus Ara\'{u}jo, Cyril Branciard, Fabio Costa, Adrien Feix, Christina Giarmatzi, and {\v{C}}aslav Brukner.
\newblock ``Witnessing causal nonseparability''.
\newblock \href{https://doi.org/10.1088/1367-2630/17/10/102001}{New J. Phys. {\bf 17}, 102001}~(2015).
\newblock  \href{http://arxiv.org/abs/1506.03776}{arXiv:1506.03776}.

\bibitem{oreshkov16}
Ognyan Oreshkov and Christina Giarmatzi.
\newblock ``Causal and causally separable processes''.
\newblock \href{https://doi.org/10.1088/1367-2630/18/9/093020}{New J. Phys. {\bf 18}, 093020}~(2016).
\newblock  \href{http://arxiv.org/abs/1506.05449}{arXiv:1506.05449}.

\bibitem{feix16}
Adrien Feix, Mateus Ara{\'u}jo, and {\v C}aslav Brukner.
\newblock ``Causally nonseparable processes admitting a causal model''.
\newblock \href{https://doi.org/10.1088/1367-2630/18/8/083040}{New J. Phys. {\bf 18}, 083040}~(2016).
\newblock  \href{http://arxiv.org/abs/1604.03391}{arXiv:1604.03391}.

\bibitem{wechs21}
Julian Wechs, Hippolyte Dourdent, Alastair~A. Abbott, and Cyril Branciard.
\newblock ``Quantum circuits with classical versus quantum control of causal order''.
\newblock \href{https://doi.org/10.1103/PRXQuantum.2.030335}{PRX Quantum {\bf 2}, 030335}~(2021).
\newblock  \href{http://arxiv.org/abs/2101.08796}{arXiv:2101.08796}.

\bibitem{purves21}
Tom Purves and Anthony~J. Short.
\newblock ``Quantum theory cannot violate a causal inequality''.
\newblock \href{https://doi.org/10.1103/PhysRevLett.127.110402}{Phys. Rev. Lett. {\bf 127}, 110402}~(2021).
\newblock  \href{http://arxiv.org/abs/2101.09107}{arXiv:2101.09107}.

\bibitem{dourdent21}
Hippolyte Dourdent, Alastair~A. Abbott, Nicolas Brunner, Ivan \ifmmode \check{S}\else \v{S}\fi{}upi\ifmmode~\acute{c}\else \'{c}\fi{}, and Cyril Branciard.
\newblock ``Semi-device-independent certification of causal nonseparability with trusted quantum inputs''.
\newblock \href{https://doi.org/10.1103/PhysRevLett.129.090402}{Phys. Rev. Lett. {\bf 129}, 090402}~(2022).
\newblock  \href{http://arxiv.org/abs/2107.10877}{arXiv:2107.10877}.

\bibitem{bowles18}
Joseph Bowles, Ivan \ifmmode \check{S}\else \v{S}\fi{}upi\ifmmode~\acute{c}\else \'{c}\fi{}, Daniel Cavalcanti, and Antonio Ac\'{\i}n.
\newblock ``Device-independent entanglement certification of all entangled states''.
\newblock \href{https://doi.org/10.1103/PhysRevLett.121.180503}{Phys. Rev. Lett. {\bf 121}, 180503}~(2018).
\newblock  \href{http://arxiv.org/abs/1801.10444}{arXiv:1801.10444}.

\bibitem{bowles18a}
Joseph Bowles, Ivan {\v{S}}upi{\'{c}}, Daniel Cavalcanti, and Antonio Ac{\'{\i}}n.
\newblock ``Self-testing of {P}auli observables for device-independent entanglement certification''.
\newblock \href{https://doi.org/10.1103/physreva.98.042336}{Phys. Rev. A {\bf 98}, 042336}~(2018).
\newblock  \href{http://arxiv.org/abs/1801.10446}{arXiv:1801.10446}.

\bibitem{davies70}
E.~B. Davies and J.~T. Lewis.
\newblock ``An operational approach to quantum probability''.
\newblock \href{https://doi.org/10.1007/BF01647093}{Commun. Math. Phys. {\bf 17}, 239--260}~(1970).

\bibitem{choi75}
Man-Duen Choi.
\newblock ``Completely positive linear maps on complex matrices''.
\newblock \href{https://doi.org/10.1016/0024-3795(75)90075-0}{Linear Algebra Appl. {\bf 10}, 285--290}~(1975).

\bibitem{chiribella08}
Giulio Chiribella, Giacomo~Mauro D'Ariano, and Paulo Perinotti.
\newblock ``Quantum circuit architecture''.
\newblock \href{https://doi.org/10.1103/PhysRevLett.101.060401}{Phys. Rev. Lett. {\bf 101}, 060401}~(2008).
\newblock  \href{http://arxiv.org/abs/0712.1325}{arXiv:0712.1325}.

\bibitem{chiribella09}
Giulio Chiribella, Giacomo~Mauro D'Ariano, and Paulo Perinotti.
\newblock ``{Theoretical framework for quantum networks}''.
\newblock \href{https://doi.org/10.1103/PhysRevA.80.022339}{Phys. Rev. A {\bf 80}, 022339}~(2009).
\newblock  \href{http://arxiv.org/abs/0904.4483}{arXiv:0904.4483}.

\bibitem{supic17}
Ivan \ifmmode \check{S}\else \v{S}\fi{}upi\ifmmode~\acute{c}\else \'{c}\fi{}, Paul Skrzypczyk, and Daniel Cavalcanti.
\newblock ``Measurement-device-independent entanglement and randomness estimation in quantum networks''.
\newblock \href{https://doi.org/10.1103/PhysRevA.95.042340}{Phys. Rev. A {\bf 95}, 042340}~(2017).
\newblock  \href{http://arxiv.org/abs/1702.04752}{arXiv:1702.04752}.

\bibitem{hoban18}
Matty~J. Hoban and Ana~Bel{\'{e}}n Sainz.
\newblock ``A channel-based framework for steering, non-locality and beyond''.
\newblock \href{https://doi.org/10.1088/1367-2630/aabea8}{New J. Phys. {\bf 20}, 053048}~(2018).
\newblock  \href{http://arxiv.org/abs/1708.00750}{arXiv:1708.00750}.

\bibitem{branciard16a}
Cyril Branciard.
\newblock ``Witnesses of causal nonseparability: an introduction and a few case studies''.
\newblock \href{https://doi.org/10.1038/srep26018}{Sci. Rep. {\bf 6}, 26018}~(2016).
\newblock  \href{http://arxiv.org/abs/1603.00043}{arXiv:1603.00043}.

\bibitem{buscemi12}
Francesco Buscemi.
\newblock ``All entangled quantum states are nonlocal''.
\newblock \href{https://doi.org/10.1103/PhysRevLett.108.200401}{Phys. Rev. Lett. {\bf 108}, 200401}~(2012).
\newblock  \href{http://arxiv.org/abs/1106.6095}{arXiv:1106.6095}.

\bibitem{branciard13}
Cyril Branciard, Denis Rosset, Yeong-Cherng Liang, and Nicolas Gisin.
\newblock ``Measurement-device-independent entanglement witnesses for all entangled quantum states''.
\newblock \href{https://doi.org/10.1103/PhysRevLett.110.060405}{Phys. Rev. Lett. {\bf 110}, 060405}~(2013).
\newblock  \href{http://arxiv.org/abs/1210.8037}{arXiv:1210.8037}.

\bibitem{mayers04}
Dominic Mayers and Andrew Yao.
\newblock ``Self testing quantum apparatus''.
\newblock \href{https://doi.org/10.26421/QIC4.4-3}{Quantum Inf. Comput. {\bf 4}, 273}~(2004).
\newblock  \href{http://arxiv.org/abs/quant-ph/0307205}{arXiv:quant-ph/0307205}.

\bibitem{supic20a}
Ivan {\v{S}}upi{\'{c}} and Joseph Bowles.
\newblock ``Self-testing of quantum systems: a review''.
\newblock \href{https://doi.org/10.22331/q-2020-09-30-337}{{Quantum} {\bf 4}, 337}~(2020).
\newblock  \href{http://arxiv.org/abs/1904.10042}{arXiv:1904.10042}.

\bibitem{rubino17}
Giulia Rubino, Lee~A. Rozema, Adrien Feix, Mateus Ara{\'u}jo, Jonas~M. Zeuner, Lorenzo~M. Procopio, {\v C}aslav Brukner, and Philip Walther.
\newblock ``Experimental verification of an indefinite causal order''.
\newblock \href{https://doi.org/10.1126/sciadv.1602589}{Sci. Adv. {\bf 3}, e1602589}~(2017).
\newblock  \href{http://arxiv.org/abs/1608.01683}{arXiv:1608.01683}.

\bibitem{goswami18}
K.~Goswami, C.~Giarmatzi, M.~Kewming, F.~Costa, C.~Branciard, J.~Romero, and A.~G. White.
\newblock ``Indefinite causal order in a quantum switch''.
\newblock \href{https://doi.org/10.1103/PhysRevLett.121.090503}{Phys. Rev. Lett. {\bf 121}, 090503}~(2018).
\newblock  \href{http://arxiv.org/abs/1803.04302}{arXiv:1803.04302}.

\bibitem{gogioso23}
Stefano Gogioso and Nicola Pinzani.
\newblock ``The geometry of causality''.
\newblock preprint~(2023) \href{http://arxiv.org/abs/2303.09017}{arXiv:2303.09017}.

\bibitem{lugt22}
Tein van~der Lugt, Jonathan Barrett, and Giulio Chiribella.
\newblock ``Device-independent certification of indefinite causal order in the quantum switch''.
\newblock \href{https://doi.org/10.1038/s41467-023-40162-8}{Nat. Commun. {\bf 14}, 5811}~(2023).
\newblock  \href{http://arxiv.org/abs/2208.00719}{arXiv:2208.00719}.

\bibitem{lugt23}
Tein van~der Lugt and Nick Ormrod.
\newblock ``Possibilistic and maximal indefinite causal order in the quantum switch''.
\newblock preprint~(2023) \href{http://arxiv.org/abs/2311.00557}{arXiv:2311.00557}.

\bibitem{kaniewski17}
J{\k e}drzej Kaniewski.
\newblock ``Self-testing of binary observables based on commutation''.
\newblock \href{https://doi.org/10.1103/PhysRevA.95.062323}{Phys. Rev. A {\bf 95}, 062323}~(2017).
\newblock  \href{http://arxiv.org/abs/1702.06845}{arXiv:1702.06845}.

\bibitem{bavaresco19}
Jessica Bavaresco, Mateus Ara{\'{u}}jo, {\v{C}}aslav Brukner, and Marco~T{\'{u}}lio Quintino.
\newblock ``Semi-device-independent certification of indefinite causal order''.
\newblock \href{https://doi.org/10.22331/q-2019-08-19-176}{{Quantum} {\bf 3}, 176}~(2019).
\newblock  \href{http://arxiv.org/abs/1903.10526}{arXiv:1903.10526}.

\end{thebibliography}

\clearpage
\onecolumngrid
\appendix


\renewcommand{\theequation}{A\arabic{equation}}
\setcounter{equation}{0}

\section{Self-testing the quantum inputs}
\label{app:st}

In this appendix we begin by recalling briefly the framework of self-testing~\cite{mayers04,supic20a}, before considering in more detail the self-testing of maximally entangled states and tomographically complete sets of orthogonal Pauli measurements, and how this can be used to self-test the remotely prepared quantum inputs in the NDI scenario.

We will focus here on Charlie and Alice, who share a quantum state $\uppsi_1^{\pC\pA}$ and perform local measurements on their share of the state described by POVMs $(\pM_{c|z}^{\pC})_c$ and $(\pM_{a|x}^{\pA})_a$, respectively.
The self-testing results for Bob and Daisy can be derived in an entirely analogous way.

The general goal of self-testing is to use the observed correlations $P(c,a|z,x)$ to show that the state $\uppsi_1^{\pC\pA}$ and/or the measurements $(\pM_{c|z}^{\pC})_c$, $(\pM_{a|x}^{\pA})_a$ must be of a specific, desired form. 
More precisely, the observed correlations are given by the Born rule as%
\footnote{Note that as we work here with the Choi representation, the measurement operators are transposed with respect to those that one would usually write (e.g., the Choi representation for a projector onto a state $\ket{\phi}$ is $(\ketbra{\phi}{\phi})^T$). This also explains the presence of transposes on $(\pM_{c|z}^\pC)^T$ and $(M_{c|z}^\rC)^T$ in the self-testing statement of Eq.~\eqref{eq:SelfTestCA} and in the subsequent equations below, compared to the notations used in the literature~\cite{bowles18,bowles18a}.}
\begin{equation}
	\label{eq:Born_rule}
    P(c,a|z,x) = (\pM_{c|z}^{\pC}\otimes \pM_{a|x}^{\pA}) * \uppsi_1^{\pC\pA} = \Tr\big[(\pM_{c|z}^{\pC}\otimes \pM_{a|x}^{\pA})^T \,  \uppsi_1^{\pC\pA}\big], \qquad \forall z,x,c,a,
\end{equation}
in a ``physical'' experiment described by $(\uppsi_1^{\pC\pA}, (\pM_{c|z}^{\pC})_c, (\pM_{a|x}^{\pA})_a)$, which we would typically like to relate (in part or wholly) to a ``reference'' experiment $(\ketbra{\psi_1}{\psi_1}^{\rC\rA},(M_{c|z}^{\rC})_c, (M_{a|x}^{\rA})_a)$.
In general, one cannot identify a unique reference experiment compatible with the observed correlations, but only an equivalence class of such experiments.
Indeed, there are two classes of transformations of quantum states and measurements which do not change the observed correlations.
The first are local isometries, encompassing local rotations and embeddings in Hilbert spaces of higher dimension. 
The other is a simultaneous transposition of the underlying quantum state and measurements.
We thus seek to show a relation between the physical and reference experiments up to such local isometries and transpositions.


\subsection*{Self-testing the maximally entangled state and Pauli measurements}

For simplicity, let us focus here on the case of $\dimd=2$ where the quantum inputs are two-dimensional.
We will discuss at the end of the section how the self-testing procedure can be generalised to arbitrary $\dimd>2$.

In order to self-test the remotely prepared quantum inputs, we will first show how Alice and Charlie can self-test the fact that they effectively share a maximally entangled state $\ket{\phi^+}^{\rC\rA}=(\ket{00}+\ket{11})/\sqrt{2}$ and that Charlie's reference measurements $(M_{c|z}^{\rC})_c$ effectively correspond to the tomographically complete Pauli measurements $\sigma_x,\sigma_y,\sigma_z$.

Although the physical state $\uppsi_1^{\pC\pA}$ that Charlie and Alice share may not be pure, it will be convenient to work with a purification $\ket{\uppsi_1}^{\pC\pA R}$ of $\uppsi_1^{\pC\pA}$, where $\HS^R$ is a purification system that the measurements $(\pM_{c|z}^{\pC})_c$ and $(\pM_{a|x}^{\pA})_a$ act, by definition, trivially upon.
Since no operations act nontrivially on this space, we will generally, in a slight abuse of notation, omit writing explicitly this space, writing simply $\ket{\uppsi_1}^{\pC\pA}$ and leave identities acting on $\HS^R$ implicit.
For simplicity, we will also assume in what follows that the physical measurements $(\pM_{c|z}^{\pC})_c$ and $(\pM_{a|x}^{\pA})_a$ are projective. Although common in self-testing proofs (and justifiable on reasonable grounds~\cite{supic20a}), this assumption is not a fundamental one in our NDI protocol, and could be removed using the techniques of~\cite{kaniewski17} to \emph{derive} this projectivity from the maximal violation of a Bell inequality.

The self-testing of a maximally entangled state and the three mutually orthogonal Pauli measurements can be obtained using an extended form of the CHSH inequality, as shown in Lemma 1 of~\cite{bowles18, bowles18a}.
Here we simply restate the elements of the results necessary for what follows, directing the interested reader to the proofs in~\cite{bowles18, bowles18a}.

Let us consider an experiment involving three measurements for Charlie and six for Alice, labelled $z=1,\dots,3$ and $x=1,\dots,6$ respectively, all of which are binary valued, and denote $\mathcal{E}_{z,x} = P(c=a|z,x) - P(c\neq a|z,x)$ the expectation value of the correlation between the outputs $c$ and $a$ for inputs $z$ and $x$.
Starting from the extended CHSH inequality
\begin{equation}\label{extCHSH}
   \mathcal{E} := \mathcal{E}_{1,1} + \mathcal{E}_{1,2} + \mathcal{E}_{2,1} - \mathcal{E}_{2,2} + \mathcal{E}_{1,3} + \mathcal{E}_{1,4} -\mathcal{E}_{3,3} + \mathcal{E}_{3,4} + \mathcal{E}_{2,5} + \mathcal{E}_{2,6} - \mathcal{E}_{3,5} + \mathcal{E}_{3,6} \leq 6
\end{equation}
obtained by the sum of three CHSH inequalities, each with a local bound of 2, let us consider the Bell inequality $\mathcal{I}= 6 - \mathcal{E}\ge 0$.
In the experiment we consider (described by quantum theory), the correlators $\mathcal{E}_{z,x}$ are obtained as
\begin{align}
	    \mathcal{E}_{z,x} &= (\mathsf{M}_{0|z}^{\pC}\otimes \mathsf{M}_{0|x}^{\pA}) * \uppsi_1^{\pC\pA} + (\mathsf{M}_{1|z}^{\pC}\otimes \mathsf{M}_{1|x}^{\pA}) * \uppsi_1^{\pC\pA} - (\mathsf{M}_{0|z}^{\pC}\otimes \mathsf{M}_{1|x}^{\pA}) * \uppsi_1^{\pC\pA} - (\mathsf{M}_{1|z}^{\pC}\otimes \mathsf{M}_{0|x}^{\pA}) * \uppsi_1^{\pC\pA}.
\end{align}
Since $\mathcal{E}$ is the sum of three CHSH inequalities, its maximum quantum value is $\mathcal{E}_Q=6\sqrt{2}$, and the maximum quantum violation of $\mathcal{I}\ge 0$ is $\mathcal{I}_Q = 6 - 6\sqrt{2} < 0$.
If this value is reached by Alice and Charlie's correlations $P_y(c,a|z,x)$ (for some input $y$ of Bob) in their physical experiment, then by Lemma~1 of Refs.~\cite{bowles18,bowles18a} there exist local isometries $V^{\mathbf{C}}: \HS^\pC\to \HS^{\rC\jC\fC}$ and $V^{\mathbf{A}}: \HS^\pA\to \HS^{\rA\jA\fA}$, where $\jC$ and $\jA$ are ``junk'' systems and $\fC$ and $\fA$ are (2-dimensional) ``flag'' systems, such that
\begin{align}
	\label{eq:SelfTestCA_stateOnly}
V^{\mathbf{C}}\otimes V^{\mathbf{A}}\ket{\uppsi_1}^{\pC\pA}&= \ket{\phi^+}^{\rC\rA}\otimes\ket{\xi_0}^{\jC\jA}\otimes\ket{00}^{\fC\fA} + \ket{\phi^+}^{\rC\rA}\otimes\ket{\xi_1}^{\jC\jA}\otimes\ket{11}^{\fC\fA}
\end{align}
and
\begin{align}
	\label{eq:SelfTestCA}
    V^{\mathbf{C}}\otimes V^{\mathbf{A}}\left((\pM_{c|z}^\pC)^T\ket{\uppsi_1}^{\pC\pA}\right) &= \left((M_{c|z}^\rC)^T\ket{\phi^+}^{\rC\rA}\right)\otimes\ket{\xi_0}^{\jC\jA}\otimes\ket{00}^{\fC\fA} + \left(M_{c|z}^\rC\ket{\phi^+}^{\rC\rA}\right)\otimes\ket{\xi_1}^{\jC\jA}\otimes\ket{11}^{\fC\fA}, 
\end{align}
where we omit identity operators acting on Alice's side, the junk states $\ket{\xi_0}^{\jC\jA}$ and $\ket{\xi_1}^{\jC\jA}$ satisfy $\|\ket{\xi_0}\|^2 + \|\ket{\xi_1}\|^2 = 1$, and Charlie's reference measurements $M_{c|z}^\rC$ are precisely the tomographically complete set of projectors 
\begin{equation}
	\label{eq:reference_measurements}
	(M_{c|1}^\rC)^T = \tfrac{1}{2}(\id + (-1)^c\sigma_Z), \quad (M_{c|2}^\rC)^T = \tfrac{1}{2}(\id + (-1)^c\sigma_X),\quad (M_{c|3}^\rC)^T = \tfrac{1}{2}(\id + (-1)^c\sigma_Y).
\end{equation}

From this self-testing result, we can determine how the remotely prepared inputs $\uprho_{c|z}^{\pA}\propto \pM_{c|z}^{\pA}*\uppsi_1^{\pC\pA}$ relate to the corresponding reference quantum inputs $\rho_{c|z}^{\rA}\propto M_{c|z}^{\rC}*\ketbra{\phi^+}{\phi^+}^{\rC\rA}$, which here are precisely the tomographically complete states $\rho_{c|z}^{\rA}=(M_{c|z}^\rC)^T$ specified in Eq.~\eqref{eq:reference_measurements}. 
Multiplying Eqs.~\eqref{eq:SelfTestCA_stateOnly} and~\eqref{eq:SelfTestCA} by ${V^{\mathbf{C}}}^\dagger\otimes {V^{\mathbf{A}}}^\dagger$, 
and recalling that $\uppsi_1^{\pC\pA}=\Tr_R \ketbra{\uppsi_1}{\uppsi_1}^{\pC\pA R}$ (with the implicit purification space made explicit here, but left implicit below on both the state and in the partial trace), it follows that (again with some implicit identity operators)
\begin{align}
    \pM_{c|z}^\pC * \uppsi_1^{\pC\pA} & = \Tr_\pC \left( (\pM_{c|z}^\pC)^T\ketbra{\uppsi_1}{\uppsi_1}^{\pC\pA} \right) \notag \\
    & = {V^{\mathbf{A}}}^\dagger\!\left(\Tr_\pC \left[ {V^{\mathbf{C}}}^\dagger \left(\sum_{k,l=0,1}\left((M_{c|z}^\rC)^{T^{\bar{k}}}\ketbra{\phi^+}{\phi^+}^{\rC\rA}\right) \otimes\ketbra{\xi_k}{\xi_l}^{\jC\jA}\otimes\ketbra{kk}{ll}^{\fC\fA}\right) V^{\mathbf{C}}\right]\right)\!V^{\mathbf{A}} \notag \\
    & = {V^{\mathbf{A}}}^\dagger\!\left(\!\Tr_{\rC\jC\fC}\! \left[ V^{\mathbf{C}}{V^{\mathbf{C}}}^\dagger \!\left(\sum_{k,l=0,1}\!\left((M_{c|z}^\rC)^{T^{\bar{k}}}\ketbra{\phi^+}{\phi^+}^{\rC\!\rA}\right) \otimes\ketbra{\xi_k}{\xi_l}^{\jC\!\jA}\otimes\ketbra{kk}{ll}^{\fC\!\fA}\right)\right]\right)\!V^{\mathbf{A}}, \label{eq:TrCMpsiM}
\end{align}
where $(\cdot)^{T^{\bar{k}}}$ denotes the ``transpose to the power $\bar{k}:=1-k$''. Noting now that $V^{\mathbf{C}}{V^{\mathbf{C}}}^\dagger$ effectively acts as the identity on the state of Eq.~\eqref{eq:SelfTestCA}, it can be removed from the previous equation. The partial trace then kills the cross terms with $k\neq l$ and we obtain
\begin{align}
    P(c|z)\, \uprho_{c|z}^\pA = \pM_{c|z}^\pC * \uppsi_1^{\pC\pA} = P(c|z)\, {V^{\mathbf{A}}}^\dagger\left( \sum_{k=0,1} (\rho_{c|z}^\rA)^{T^k}\otimes\xi_k^{\jfA} \right) V^{\mathbf{A}}, \label{eq:MC_rhoCA}
\end{align}
with $\jfA = \jA\fA$, $\xi_k^{\jfA} = \Tr_{\jC}\ketbra{\xi_k}{\xi_k}^{\jC\jA}\otimes\ketbra{k}{k}^{\fA}$ such that $\sum_k \Tr(\xi_k^{\jfA}) = 1$, and $P(c|z)\, \rho_{c|z}^\rA = \Tr_{\rC} \left((M_{c|z}^\rC)^T\ketbra{\phi^+}{\phi^+}^{\rC\rA}\right) = M_{c|z}^\rC * \ketbra{\phi^+}{\phi^+}^{\rC\rA}$.%
\footnote{One may note that up to a proportionality constant, $\rho_{c|z}^\rA$ is formally the same matrix as $M_{c|z}^\rC$, in the space $\rA$ rather than $\rC$. (This can indeed be interpreted as some form of quantum teleportation from system $\rC$ to system $\rA$.)}

By the symmetry of the considered network scenario (cf.\ Eq.~\eqref{eq:Pabcd_xyzw_NDI} and Fig.~\ref{fig:scenarii}), we obtain a similar self-testing statement for Bob and Daisy:  provided that their correlations $P_x(b,d|y,w)$ also violate the Bell inequality Eq.~\eqref{extCHSH} up to its quantum maximum (for some input $x$ of Alice), there exists a local isometry of the form $V^{\mathbf{B}}: \HS^\pB\to \HS^{\rB\jfB}$ such that we can write
\begin{align}
    P(d|w)\, \uprho_{d|w}^\pB = \pM_{d|w}^\pD * \uppsi_2^{\pB\pD} = P(d|w)\, {V^{\mathbf{B}}}^\dagger\left( \sum_{l=0,1} (\rho_{d|w}^\rB)^{T^l}\otimes\zeta_l^{\jfB} \right) V^{\mathbf{B}}, \label{eq:MD_rhoBD}
\end{align}
with $\sum_l \Tr(\zeta_l^{\jfB}) = 1$ and $P(d|w)\, \rho_{d|w}^\rB = \Tr_{\rD} \left((M_{d|w}^\rD)^T\ketbra{\phi^+}{\phi^+}^{\rB\rD}\right) = M_{d|w}^\rD * \ketbra{\phi^+}{\phi^+}^{\rB\rD}$.

Thus, we have shown that both remotely prepared inputs $\uprho_{c|z}^\pA\propto \pM_{c|z}^\pC * \uppsi_1^{\pC\pA}$ and $\uprho_{d|w}^\pB\propto \pM_{d|w}^\pD * \uppsi_2^{\pB\pD}$ can be related to the reference quantum inputs $\rho_{c|z}^\rA\propto M_{c|z}^\rC * \ketbra{\phi^+}{\phi^+}^{\rC\rA}$ and $\rho_{d|w}^\rB\propto M_{d|w}^\rD*\ketbra{\phi^+}{\phi^+}^{\rB\rD}$ via local isometries.
That is, we are able to turn the self-test of the maximally entangled state that Charlie and Alice share, along with Charlie's tomographically complete set of orthogonal Pauli measurements, into a self-test of the quantum inputs used in the reference NSDI-QI certification protocol (and likewise for Bob and Daisy).
Note that in the NDI protocol presented in the main text, we assumed that Charlie's (tomographically complete) quantum inputs in the reference scenario are obtained as the remotely prepared states when $c=0$ for the different inputs $z$, whereas here the $c=0$ outcomes only give one projector for each measurement, which is not quite a tomographically complete set. 
This detail can be overcome by considering some additional measurements in the NDI protocol obtained by relabelling some of the above measurements' outcomes, so as to complete the set of reference quantum inputs (and likewise for Daisy's measurements).

Finally, while the above approach, presented explicitly for the case of two-dimensional quantum inputs (i.e., $\dimd=2$), it can be readily generalised to $\dimd=2^m$, for $m> 1$.
Indeed, the maximally entangled state $\ket{\phi^+_\dimd}^{\rC\rA}=\frac{1}{\sqrt{\dimd}}\sum_{j=0}^{\dimd-1}\ket{j,j}$, for $\dimd=2^m$, can be self-tested through the parallel repetition of the extended CHSH inequality~\eqref{extCHSH}~\cite{bowles18a}.
In this case, Charlie has $3^m$ measurement settings corresponding to the nontrivial tensor products of orthogonal single-qubit Pauli measurements, each of which has $2^m$ outcomes so that the $M_{c|z}^\rC$ one self-tests are precisely the projectors onto the $6^m$ eigenstates of $m$-fold Pauli measurements. 
As in the $\dimd=2$ case, the quantum inputs prepared on the $c=0$ outcomes can be extended to a tomographically complete set by adding some additional measurements with relabelled outcomes.

For $\dimd\,\neq 2^m$, the above procedure can again be utilised by first embedding the reference scenario into some $\dimd\,'$-dimensional spaces for some $\dimd\,'=2^m > \dimd$.
By writing the D-POVM $\mathbb{E}^{\rA\rB}$ in these $\dimd\,'$-dimensional spaces (and taking $\HS^{\rC}$ and $\HS^{\rD}$ to hence also be $\dimd\,'$-dimensional) one can then proceed using the general approach described above for $\dimd\,'=2^m$ before restricting ourselves to quantum inputs in the $\dimd$-dimensional subspace of interest. 


\section{Certifying causal nonseparability in a network-device-independent manner}\label{app:CNSep_NDI}

\renewcommand{\theequation}{B\arabic{equation}}
\setcounter{equation}{0}

The self-testing statements above allow us to directly relate the observed correlations to those of the ideal reference systems in the NSDI-QI setup.
To see this, let us define for convenience $\dket{{V^{\mathbf{A}}}^\dagger}^{\rA\jfA\pA} \coloneqq \sum_i \ket{i}^{\rA\jfA} \otimes {V^{\mathbf{A}}}^\dagger\ket{i}^{\rA\jfA}$, where $\{\ket{i}^{\rA\jfA}\}_i$ is the computational basis of $\HS^{\rA\jfA}$. With these notations one can write, using the link product,
\begin{align}
    {V^{\mathbf{A}}}^\dagger\big[ (\rho_{c|z}^\rA)^{T^k}\otimes\xi_k^{\jfA} \big] V^{\mathbf{A}} & = \dketbra{{V^{\mathbf{A}}}^\dagger}{{V^{\mathbf{A}}}^\dagger}^{\rA\jfA\pA} * \big[ (\rho_{c|z}^\rA)^{T^k}\otimes\xi_k^{\jfA} \big] = \Big( \dketbra{{V^{\mathbf{A}}}^\dagger}{{V^{\mathbf{A}}}^\dagger}^{\rA\jfA\pA} * \xi_k^{\jfA} \Big)^{T_{\rA}^k} * \rho_{c|z}^\rA, 
\end{align}
and similarly for ${V^{\mathbf{B}}}^\dagger\big[ (\rho_{d|w}^\rB)^{T^l}\otimes\zeta_l^{\jfB} \big] V^{\mathbf{B}}$, introducing $\dket{{V^{\mathbf{B}}}^\dagger}^{\rB\jfB\pB}$ in a similar way to $\dket{{V^{\mathbf{A}}}^\dagger}^{\rA\jfA\pA}$ above.
From Eqs.~\eqref{eq:MC_rhoCA} and~\eqref{eq:MD_rhoBD}, the correlations $P(c,a,b,d|z,x,y,w)$ can then be written as
\begin{align}
& P(c,a,b,d|z,x,y,w) = \left(\pM_{c|z}^\pC \otimes \pM_{a|x}^{\pA A}\otimes \pM_{b|y}^{\pB B} \otimes \pM_{d|w}^\pD\right) * \left(\uppsi_1^{\pC\pA}\otimes W^{AB} \otimes \uppsi_2^{\pB\pD}\right) \notag \\[1mm]
& \quad = P(c|z)\, P(d|w) \left( \pM_{a|x}^{\pA A}\otimes \pM_{b|y}^{\pB B} \right) * W^{AB} * \Bigg[ \Bigg( \sum_{k=0,1} {V^{\mathbf{A}}}^\dagger\big[ (\rho_{c|z}^\rA)^{T^k}\otimes\xi_k^{\jfA} \big] V^{\mathbf{A}} \Bigg) \otimes \Bigg( \sum_{l=0,1} {V^{\mathbf{B}}}^\dagger \big[ (\rho_{d|w}^\rB)^{T^l}\otimes\zeta_l^{\jfB} \big] V^{\mathbf{B}} \Bigg) \Bigg] \notag \\[1mm]
& \quad = P(c|z)\, P(d|w) \, \mathsf{E}_{a,b|x,y}^{\pA\pB} * \!\Bigg[\! \Bigg(\! \sum_{k=0,1} \!\!\Big(\! \dketbra{{V^{\mathbf{A}}}^\dagger}{{V^{\mathbf{A}}}^\dagger}^{\rA\jfA\pA} \!* \xi_k^{\jfA} \Big)^{\!T_{\rA}^k} \!\!* \rho_{c|z}^\rA \!\Bigg)  \!\otimes\! \Bigg(\! \sum_{l=0,1} \!\!\Big(\! \dketbra{{V^{\mathbf{B}}}^\dagger}{{V^{\mathbf{B}}}^\dagger}^{\rB\jfB\pB} \!* \zeta_l^{\jfB} \Big)^{\!T_{\rB}^l} \!\!* \rho_{d|w}^\rB \!\Bigg) \!\Bigg]\! \notag \\[1mm]
& \quad = P(c|z)\, P(d|w)\ E_{a,b|x,y}^{\rA\rB} * \left( \rho_{c|z}^\rA \otimes \rho_{d|w}^\rB \right) \label{eq:P_cabd_zxyw}
\end{align}
with the ``physical'' D-POVM elements $\mathsf{E}_{a,b|x,y}^{\pA\pB}=\big(\mathsf{M}_{a|x}^{\pA A}\otimes \mathsf{M}_{b|y}^{\pB B}\big)*W^{AB}$ and the new D-POVM elements effectively acting on the ``reference'' quantum input states
\begin{align}
E_{a,b|x,y}^{\rA\rB} & \coloneqq \mathsf{E}_{a,b|x,y}^{\pA\pB} * \Bigg[ \sum_{k=0,1} \left( \dketbra{{V^{\mathbf{A}}}^\dagger}{{V^{\mathbf{A}}}^\dagger}^{\rA\jfA\pA} * \xi_k^{\jfA} \right)^{T_{\rA}^k} \otimes \sum_{l=0,1} \left( \dketbra{{V^{\mathbf{B}}}^\dagger}{{V^{\mathbf{B}}}^\dagger}^{\rB\jfB\pB} * \zeta_l^{\jfB} \right)^{T_{\rB}^l} \Bigg]. \label{eq:rE_from_pE}
\end{align}

The correlations (cf.\ Eq.~\eqref{eq:def_Pstar})
\begin{equation}
	\label{eq:effectiveDPOVMcorrs}
	P_\star(a,b|z,w) = P(a,b|z, x{=}\star, y{=}\star, w, c{=}0, d{=}0) = E_{a,b|\star,\star}^{\rA\rB} * \left( \rho_z^\rA \otimes \rho_w^\rB \right) = P'(a,b|\rho_z^\rA,\rho_w^\rB)
\end{equation}
thus describe precisely the correlations in a SDI-QI setting with the reference quantum inputs $\rho_z^\rA$ and $\rho_w^\rB$ and the effective D-POVM $\mathbb{E}_\text{eff}^{\rA\rB}=(E_{a,b|*,*}^{\rA\rB})_{a,b}$ involving the D-POVM elements defined in Eq.~\eqref{eq:rE_from_pE}.

We now show that the local transformations used in this self-testing procedure preserve causal separability, thus allowing us to certify the causal nonseparability of the physical D-POVM $\mathbb{E}^{\pA\pB}=(\mathsf{E}_{a,b|*,*}^{\pA\pB})_{a,b}$---and hence the underlying physical process $W^{AB}$---despite lacking any knowledge of the Hilbert spaces upon which it acts.

Let us for that assume that the process matrix is compatible with the causal order $A\prec B$. 
Such a process matrix takes the form $W = W^{A\prec B_I}\otimes \id^{B_O}$, with $\Tr_{B_I} W^{A\prec B_I} = W^{A_I}\otimes \id^{A_O}$ and $\Tr W^{A_I} = 1$. Using the fact that $\sum_b \pM_{b|y}^{\pB B} * \id^{B_O} = \sum_b \Tr_{B_O}\pM_{b|y}^{\pB B} = \id^{\pB B_I}$ according to the overall trace-preserving condition for a quantum instrument, we have
\begin{align}
    \sum_b \mathsf{E}_{a,b|x,y}^{\pA\pB} = \sum_b \big(\mathsf{M}_{a|x}^{\pA A}\otimes \mathsf{M}_{b|y}^{\pB B}\big) * W^{A\prec B_I}\otimes \id^{B_O} = \big(\mathsf{M}_{a|x}^{\pA A}\otimes \id^{\pB B_I} \big) * W^{A\prec B_I} = \mathsf{E}_{a|x}^{\pA} \otimes \id^{\pB}
\end{align}
with $\mathsf{E}_{a|x}^{\pA} \coloneqq \mathsf{M}_{a|x}^{\pA A} * \Tr_{B_I} W^{A\prec B_I} = \Tr_{A_O} \mathsf{M}_{a|x}^{\pA A} * W^{A_I}$ (such that $(\mathsf{E}_{a|x}^{\pA})_a$ forms Alice's marginal POVM applied to her physical quantum input). From Eq.~\eqref{eq:rE_from_pE}, we then obtain%
\footnote{Here we use the fact that $\Tr_{\pB} \dketbra{{V^{\mathbf{B}}}^\dagger}{{V^{\mathbf{B}}}^\dagger}^{\rB\jfB\pB} = (V^{\mathbf{B}}{V^{\mathbf{B}}}^\dagger)^T$, so that $\Tr_{\pB} \dketbra{{V^{\mathbf{B}}}^\dagger}{{V^{\mathbf{B}}}^\dagger}^{\rB\jfB\pB} * \zeta_l^{\jfB} = \Tr_{\jfB} \big( V^{\mathbf{B}}{V^{\mathbf{B}}}^\dagger (\id^{\rB}\otimes \zeta_l^{\jfB}) \big)^T$. To evaluate this further, note that $V^{\mathbf{A}}{V^{\mathbf{A}}}^\dagger$ acts as the identity on the state of Eq.~\eqref{eq:SelfTestCA}, and therefore on the state obtained after tracing out over $\rC\jfC$---that is, on $\sum_k (\rho_{c|z}^\rA)^{T^k}\otimes\xi_k^{\jfA}$, for some tomographically complete set of states $\rho_{c|z}^\rA$. The same holds on Bob's side, so that $V^{\mathbf{B}}{V^{\mathbf{B}}}^\dagger [\sum_l(\rho^{\rB})^{T^{\bar{l}}}\otimes \zeta_l^{\jfB}] = \sum_l(\rho^{\rB})^{T^{\bar{l}}}\otimes \zeta_l^{\jfB}$ for all $\rho^{\rB}$. It then follows that $\Tr\big[\big( \sum_l \big( \Tr_{\pB} \dketbra{{V^{\mathbf{B}}}^\dagger}{{V^{\mathbf{B}}}^\dagger}^{\rB\jfB\pB} * \zeta_l^{\jfB} \big)^{T^l} \big)\rho^{\rB}\big] = \Tr\big[ \sum_l \Tr_{\jfB} \big( V^{\mathbf{B}}{V^{\mathbf{B}}}^\dagger (\id^{\rB}\otimes \zeta_l^{\jfB}) \big)^{T^{\bar{l}}} \rho^{\rB}\big] = \Tr\big[ V^{\mathbf{B}}{V^{\mathbf{B}}}^\dagger [\sum_l (\rho^{\rB})^{T^{\bar{l}}}\otimes \zeta_l^{\jfB}] \big] = \Tr\big[ \sum_l (\rho^{\rB})^{T^{\bar{l}}}\otimes \zeta_l^{\jfB} \big] = (\Tr \rho^{\rB}) (\sum_l \Tr \zeta_l^{\jfB}) = \Tr \rho^{\rB}$ for all $\rho^{\rB}$, which implies that $\sum_l \big( \Tr_{\pB} \dketbra{{V^{\mathbf{B}}}^\dagger}{{V^{\mathbf{B}}}^\dagger}^{\rB\jfB\pB} * \zeta_l^{\jfB} \big)^{T^l} = \id^{\rB}$. \label{ftn:sum_trB_V_V_zeta}}%
\begin{align}
\sum_b E_{a,b|x,y}^{\rA\rB} & = \mathsf{E}_{a|x}^{\pA} \otimes \id^{\pB} * \Bigg[ \sum_{k=0,1} \left( \dketbra{{V^{\mathbf{A}}}^\dagger}{{V^{\mathbf{A}}}^\dagger}^{\rA\jfA\pA} * \xi_k^{\jfA} \right)^{T_{\rA}^k} \otimes \sum_{l=0,1} \left( \dketbra{{V^{\mathbf{B}}}^\dagger}{{V^{\mathbf{B}}}^\dagger}^{\rB\jfB\pB} * \zeta_l^{\jfB} \right)^{T_{\rB}^l} \Bigg] \notag \\
 & = \Bigg[ \mathsf{E}_{a|x}^{\pA} * \sum_{k=0,1} \left( \dketbra{{V^{\mathbf{A}}}^\dagger}{{V^{\mathbf{A}}}^\dagger}^{\rA\jfA\pA} * \xi_k^{\jfA} \right)^{T_{\rA}^k} \Bigg] \otimes \sum_{l=0,1} \left( \Tr_{\pB} \dketbra{{V^{\mathbf{B}}}^\dagger}{{V^{\mathbf{B}}}^\dagger}^{\rB\jfB\pB} * \zeta_l^{\jfB} \right)^{T^l} \notag \\
 & = E_{a|x}^{\rA} \otimes \id^{\rB}. \label{eq:marginal_ErArB}
\end{align}
with
\begin{align}
E_{a|x}^{\rA} & \coloneqq \mathsf{E}_{a|x}^{\pA} * \sum_{k=0,1} \left( \dketbra{{V^{\mathbf{A}}}^\dagger}{{V^{\mathbf{A}}}^\dagger}^{\rA\jfA\pA} * \xi_k^{\jfA} \right)^{T_{\rA}^k},
\end{align}
which now forms a single-partite POVM acting on Alice's reference quantum input (one can in particular check that $\sum_a \mathsf{E}_{a|x}^{\pA} = \id^{\pA}$ and $\sum_a E_{a|x}^{\rA} = \id^{\rA}$).

Thus, from Eq.~\eqref{eq:marginal_ErArB} we can conclude that, starting from a causally ordered $W = W^{A\prec B_I}\otimes \id^{B_O}$ compatible with $A\prec B$, we obtain a D-POVM $(E_{a,b|x,y}^{\rA\rB})_{a,b}$ compatible with the order $\rA\prec\rB$. Similarly, from a process matrix compatible with the causal order $B\prec A$ we obtain a D-POVM $(E_{a,b|x,y}^{\rA\rB})_{a,b}$ compatible with the order $\rB\prec\rA$. We can then conclude that any D-POVM $(E_{a,b|x,y}^{\rA\rB})_{a,b}$ induced by a causally separable process matrix is causally separable, as defined in~\cite{dourdent21}.

This is the case, in particular, when $x=y=\star$, so that the effective D-POVM $\mathbb{E}_\text{eff}^{\rA\rB}=(E_{a,b|*,*}^{\rA\rB})_{a,b}$ is causally separable whenever $W^{AB}$ is causally separable.
As a result, the correlations $P'(a,b|\rho_z^\rA,\rho_w^\rB)$ defined in Eq.~\eqref{eq:effectiveDPOVMcorrs} obtained from a causally separable process $W^{AB}$ must satisfy any witness inequality $\mathcal{J}\ge 0$.
If these correlations violate such an inequality, giving $\mathcal{J}<0$, then this therefore proves the causal nonseparability of the process in an NDI manner.

Such a violation---and hence NDI certification---can moreover be obtained for any process $W^{AB}$ that can (for appropriately chosen reference instruments $(M_a^{\rA A})_a$ and $(M_b^{\rB B})_b$) induce a causally nonseparable D-POVM $\mathbb{E}^{\rA\rB}=(E_{a,b}^{\rA\rB})_{a,b}$ in the reference scenario, i.e., which cannot be decomposed according to Eqs.~\eqref{eq:csep_DPOVM}--\eqref{eq:dpovmba}.
Indeed, as outlined in~\cite{dourdent21}, the causal nonseparability of $\mathbb{E}^{\rA\rB}$ can always be witnessed by a family $\mathbb{S}^{\rA\rB}=(S_{a,b}^{\rA\rB})_{a,b}$ of operators (a ``witness'') such that $\sum_{a,b} S_{a,b}^{\rA\rB} * E_{a,b}^{\rA\rB} < 0$ only if $\mathbb{E}^{\rA\rB}$ is causally nonseparable.
Concretely, writing $S_{a,b}^{\rA\rB}=\sum_{z,w} s_{a,b}^{(z,w)}\rho_z^\rA \otimes \rho_w^\rB$ using the tomographically complete sets of states from the NDI-protocol, one obtains the witness inequality
\begin{equation}\label{witness}
	\mathcal{J}:=\sum_{a,b,z,w} s_{a,b}^{(z,w)} P'(a,b|\rho_z^\rA,\rho_w^\rB) \ge 0.
\end{equation}
By implementing precisely the reference NSDI-QI protocol, one simply has $E_{a,b}^{\rA\rB}=E_{a,b|*,*}^{\rA\rB}=\pE_{a,b|*,*}^{\pA\pB}$ for all $a,b$ so that the reference, effective, and physical D-POVMs all coincide, and the witness inequality can indeed, as expected, be violated.


\section{Robustness of the procedure}
\label{app:robustness}

\renewcommand{\theequation}{C\arabic{equation}}
\setcounter{equation}{0}

The exact self-testing conditions (i.e., strictly reaching the maximal quantum violation of the Bell inequality under consideration) are never met in any realistic scenario. Thus, to make self-testing applicable in practice one needs to ensure that it is ``robust''. In particular, the reference and physical experiments should be sufficiently close if the self-testing correlations are approximately reproduced by the physical experiment. 
In this section we discuss how our NDI certification procedure is affected if the states shared by Alice and Charlie and by Bob and Daisy approximately reproduce the self-testing correlations.

The robust self-testing of the maximally entangled pair of qubits (resp., qudits) is discussed in Appendix~B (resp., E) of~\cite{bowles18a}. 
Let us restrict ourselves again here, for simplicity, to the case of qubits (for qudits one can adapt the approach of~\cite{bowles18a} in a similar manner). 
The robust self-testing statements---i.e., the robust version of Eq.~\eqref{eq:SelfTestCA}---then take the form%
\footnote{Recall that we assume, for simplicity, that the physical measurements $(\pM_{c|z}^{\pC})_c$ and $(\pM_{a|x}^{\pA})_a$ are projective. In the noisy case this projectivity property cannot be derived directly by using the techniques of~\cite{kaniewski17}, which require the maximal violation of a Bell inequality, although we expect that the noise-robust statements of~\cite{kaniewski17} should still allow one to evade the strict requirement for the projectivity assumption (working out the details is left as an open question).}
\begin{multline}\label{robST}
     \bigg\|V^\mathbf{C}\otimes V^\mathbf{A} \left(({\pM}_{c|z}^\pC)^T\ket{\uppsi_1}^{\pC\pA}\right)  \\ - \bigg( \left((M_{c|z}^T)^\rC\ket{\phi^+}^{\rC\rA}\right)\otimes\ket{\xi_0}^{\jC\jA}\otimes\ket{00}^{\fC\fA} + \left(M_{c|z}^\rC\ket{\phi^+}^{\rC\rA}\right)\otimes\ket{\xi_1}^{\jC\jA}\otimes\ket{11}^{\fC\fA} \bigg) \bigg\| \leq \theta,
\end{multline}
where $\theta$ depends on the distance between observed and reference correlations (e.g., in the $\dimd = 2$ case, if the violation of the Bell inequality~\eqref{extCHSH} reaches some value $6\sqrt{2} - \epsilon$, one can take $\theta = \kappa \sqrt{\epsilon}$, for some multiplicative factor $\kappa$~\cite{bowles18a}).
Denoting by $\ket{\Psi_{c|z}}$ the state in the right-hand side of Eq.~\eqref{eq:SelfTestCA} (in the noiseless case), this equation can be rewritten as
\begin{align}\label{robSTAC}
         V^{\mathbf{C}}\otimes V^{\mathbf{A}}\left((\pM_{c|z}^\pC)^T\ket{\uppsi_1}^{\pC\pA}\right) =  \ket{\Psi_{c|z}}^{\rC\jfC\rA\jfA}+\ket{\omega_{c|z}}^{\rC\jfC\rA\jfA},
\end{align}
where $\ket{\omega_{c|z}}$ is an unknown state with vector norm smaller than or equal to $\theta$.
Eq.~\eqref{eq:MC_rhoCA} then becomes, in the noisy case,
\begin{align}
    P(c|z)\, \uprho_{c|z}^\pA = \pM_{c|z}^\pC * \uppsi_1^{\pC\pA} & = {V^{\mathbf{A}}}^\dagger \Tr_{\rC\jfC}\left( (\ket{\Psi_{c|z}}+\ket{\omega_{c|z}})(\bra{\Psi_{c|z}}+\bra{\omega_{c|z}}) \right) V^{\mathbf{A}} \notag \\
    & = P(c|z)\, {V^{\mathbf{A}}}^\dagger\left( \sum_{k=0,1} (\rho_{c|z}^\rA)^{T^k}\otimes\xi_k^{\jfA} \right) V^{\mathbf{A}} + \delta_{c|z}^{\pA}, \label{eq:M_psi1_noisy}
\end{align}
with
\begin{align}
    \delta_{c|z}^{\pA} = {V^{\mathbf{A}}}^\dagger \Tr_{\rC\jfC}\left( \ketbra{\Psi_{c|z}}{\omega_{c|z}}+\ketbra{\omega_{c|z}}{\Psi_{c|z}}+\ketbra{\omega_{c|z}}{\omega_{c|z}} \right) V^{\mathbf{A}}. \label{eq:def_delta_cz}
\end{align}
The norm of this correction term can be bounded in the same way as it is done in Appendix~G of~\cite{bowles18a}. Namely, using the triangle inequality and noting that $\|\ket{\Psi_{c|z}}\| = \frac{1}{\sqrt{2}}$, we get
\begin{equation}
   \left\| \delta_{c|z}^{\pA} \right\|_1 \leq \sqrt{2}\,\theta + \theta^2, \label{eq:UB_norm_delta}
\end{equation}
where $\|\cdot\|_1$ is the trace norm.

Similar statements can be made on Bob and Daisy's side, allowing one to write $\pM_{d|w}^\pD * \uppsi_2^{\pB\pD}$ in a similar way to Eq.~\eqref{eq:M_psi1_noisy}, in terms of a correction term $\delta_{d|w}^{\pB}$ whose trace norm is also upper-bounded as in Eq.~\eqref{eq:UB_norm_delta} (for simplicity we take the same value of $\theta$ on both sides).
The observed correlations $P(c,a,b,d|z,x,y,w)$, following Eq.~\eqref{eq:P_cabd_zxyw}, then become
\begin{align}
P(c,a,b,d|z,x,y,w) & = \mathsf{E}_{a,b|x,y}^{\pA\pB} * \left(\big(\pM_{c|z}^\pC * \uppsi_1^{\pC\pA}\big) \otimes \big(\pM_{d|w}^\pD * \uppsi_2^{\pB\pD}\big)\right) \notag \\[1mm]
& = \mathsf{E}_{a,b|x,y}^{\pA\pB} * \bigg(\! P(c|z)\sum_{k=0,1} \Big( \dketbra{{V^{\mathbf{A}}}^\dagger}{{V^{\mathbf{A}}}^\dagger}^{\rA\jfA\pA} \! * \xi_k^{\jfA} \Big)^{T_{\rA}^k} * \rho_{c|z}^\rA + \delta_{c|z}^{\pA} \bigg) \notag \\[-2mm]
& \hspace{30mm} \otimes \bigg(\! P(d|w)\sum_{l=0,1} \Big( \dketbra{{V^{\mathbf{B}}}^\dagger}{{V^{\mathbf{B}}}^\dagger}^{\rB\jfB\pB} \! * \zeta_l^{\jfB} \Big)^{T_{\rB}^l} * \rho_{d|w}^\rB + \delta_{d|w}^{\pB} \bigg) \notag \\[1mm]
& = P(c|z)\, P(d|w)\ E_{a,b|x,y}^{\rA\rB} * \left( \rho_{c|z}^\rA \otimes \rho_{d|w}^\rB \right) + P(c|z)\, \mathsf{E}_{a,b|x,y}^{\rA\pB} * \left( \rho_{c|z}^\rA \otimes \delta_{d|w}^{\pB} \right) \notag \\
& \hspace{15mm} + P(d|w)\, \mathsf{E}_{a,b|x,y}^{\pA\rB} * \left( \delta_{c|z}^{\pA} \otimes \rho_{d|w}^\rB \right) + \mathsf{E}_{a,b|x,y}^{\pA\pB} * \left( \delta_{c|z}^{\pA} \otimes \delta_{d|w}^{\pB} \right) \label{eq:P_cabd_zxyw_noisy}
\end{align}
with the ``physical'' D-POVM elements defined as $\mathsf{E}_{a,b|x,y}^{\pA\pB}=\big(\mathsf{M}_{a|x}^{\pA A}\otimes \mathsf{M}_{b|y}^{\pB B}\big)*W^{AB}$, the D-POVM elements $E_{a,b|x,y}^{\rA\rB}$ acting on the ``reference'' input states as given in Eq.~\eqref{eq:rE_from_pE}, and with the new D-POVM elements effectively acting on the ``reference'' and the ``physical'' quantum input states
\begin{align}
\mathsf{E}_{a,b|x,y}^{\rA\pB} & \coloneqq \mathsf{E}_{a,b|x,y}^{\pA\pB} * \sum_{k=0,1} \Big( \dketbra{{V^{\mathbf{A}}}^\dagger}{{V^{\mathbf{A}}}^\dagger}^{\rA\jfA\pA} \! * \xi_k^{\jfA} \Big)^{T_{\rA}^k}, \notag \\
\mathsf{E}_{a,b|x,y}^{\pA\rB} & \coloneqq \mathsf{E}_{a,b|x,y}^{\pA\pB} * \sum_{l=0,1} \Big( \dketbra{{V^{\mathbf{B}}}^\dagger}{{V^{\mathbf{B}}}^\dagger}^{\rB\jfB\pB} \! * \zeta_l^{\jfB} \Big)^{T_{\rB}^l}.
\end{align}

The correlations $P_\star(a,b|z,w)$ used in our protocol to certify causal nonseparability are then
\begin{align}
P_\star(a,b|z,w)&  = P(a,b|z, x{=}\star, y{=}\star, w, c{=}0, d{=}0) \notag \\
& = P'(a,b|\rho_z^\rA,\rho_w^\rB) + \mathsf{E}_{a,b|\star,\star}^{\rA\pB} * \left( \rho_z^\rA \otimes \frac{\delta_{0|w}^{\pB}}{P(0|w)} \right) \notag \\
& \hspace{15mm} + \mathsf{E}_{a,b|\star,\star}^{\pA\rB} * \left( \frac{\delta_{0|z}^{\pA}}{P(0|z)} \otimes \rho_w^\rB \right) + \mathsf{E}_{a,b|\star,\star}^{\pA\pB} * \left( \frac{\delta_{0|z}^{\pA}}{P(0|z)} \otimes \frac{\delta_{0|w}^{\pB}}{P(0|w)} \right), \label{eq:robustprobabilities}
\end{align} 
with $P'(a,b|\rho_z^\rA,\rho_w^\rB) = E_{a,b|\star,\star}^{\rA\rB} * \left( \rho_z^\rA \otimes \rho_w^\rB \right)$.
We can bound the second term in the right-hand side of Eq.~\eqref{eq:robustprobabilities} as
\begin{align}
    \notag \left|\Tr\left[\left(\mathsf{E}_{a,b|\star,\star}^{\rA\pB}\right)^T \left( \rho_z^\rA \otimes \frac{\delta_{0|w}^{\pB}}{P(0|w)} \right)\right] \right| & \leq \left\| \left(\mathsf{E}_{a,b|\star,\star}^{\rA\pB} \right)^T \left( \rho_z^\rA \otimes \frac{\delta_{0|w}^{\pB}}{P(0|w)} \right)\right\|_1 \leq \left\| \mathsf{E}_{a,b|\star,\star}^{\rA\pB}\right\|_\infty \left\|\rho_z^\rA \otimes \frac{\delta_{0|w}^{\pB}}{P(0|w)}\right\|_1 \\
    & \leq \frac{1}{P(0|w)} \left\|\rho_z^\rA\right\|_1 \left\| \delta_{0|w}^{\pB} \right\|_1 \leq \frac{1}{P(0|w)}\left(\sqrt{2}\,\theta + \theta^2\right), \label{robust1stterm}
\end{align}
where in the first inequality we used the fact that $\left|\Tr(A)\right| \leq  \Tr\left|A\right| = \|A\|_1$, the second one follows from H\"{o}lder's inequality, in the third inequality we used the fact that the maximal eigenvalue of $\mathsf{E}_{a,b|\star,\star}^{\rA\pB}$ cannot be larger than 1, and in the last inequality we used the normalisation of $\rho_z^\rA$ together with the analogue of the bound~\eqref{eq:UB_norm_delta} above, for $\left\| \delta_{0|w}^{\pB} \right\|_1$.
In a similar way, for the last two terms in Eq.~\eqref{eq:robustprobabilities}, we can write
\begin{align}\label{robust2ndterm}
\left|\Tr\left[\left(\mathsf{E}_{a,b|\star,\star}^{\pA\rB}\right)^T \left( \frac{\delta_{0|z}^{\pA}}{P(0|z)}  \otimes \rho_w^{B'} \right)\right] \right| \leq   \frac{1}{P(0|z)}\left(\sqrt{2}\,\theta + \theta^2\right),
\end{align}
and
\begin{align}\label{robust3rdterm}
    \left|\Tr\left[\left(\mathsf{E}_{a,b|\star,\star}^{\pA\pB}\right)^T  \left( \frac{\delta_{0|z}^{\pA}}{P(0|z)} \otimes \frac{\delta_{0|w}^{\pB}}{P(0|w)} \right)\right]\right| \leq \frac{1}{P(0|w)P(0|z)}\left(\sqrt{2}\,\theta + \theta^2\right)^2.
\end{align}

Consider now a witness inequality of the form of Eq.~\eqref{witness}, $\mathcal{J}_{\text{ref.}}:=\sum_{a,b,z,w} s_{a,b}^{(z,w)} P'(a,b|\rho_z^\rA,\rho_w^\rB) \ge 0$. Evaluating the same witness expression with the experimentally obtained probabilities $P_\star(a,b|z,w)$ instead of $P'(a,b|\rho_z^\rA,\rho_w^\rB)$, so as to define $\mathcal{J}_{\text{exp.}}:=\sum_{a,b,z,w} s_{a,b}^{(z,w)} P_\star(a,b|z,w)$, we can write
\begin{align}
    \mathcal{J}_{\text{exp.}} = \mathcal{J}_{\text{ref.}} + \mathcal{J}_{\text{correction}}
\end{align}
with
\begin{align}
    \mathcal{J}_{\text{correction}} = \sum_{a,b,z,w} s_{a,b}^{(z,w)} \left\{ \mathsf{E}_{a,b|\star,\star}^{\rA\pB} * \left( \rho_z^\rA \otimes \frac{\delta_{0|w}^{\pB}}{P(0|w)} \right) + \mathsf{E}_{a,b|\star,\star}^{\pA\rB} * \left( \frac{\delta_{0|z}^{\pA}}{P(0|z)} \otimes \rho_w^\rB \right) + \mathsf{E}_{a,b|\star,\star}^{\pA\pB} * \left( \frac{\delta_{0|z}^{\pA}}{P(0|z)} \otimes \frac{\delta_{0|w}^{\pB}}{P(0|w)} \right) \right\}.
\end{align}
The previous analysis allows us to bound $\mathcal{J}_{\text{correction}}$ as $\left|\mathcal{J}_{\text{correction}}\right| \le \mathcal{J}_{\text{correction}}^{\text{max}}$, with an upper bound $\mathcal{J}_{\text{correction}}^{\text{max}}$ on the maximum correction that can be evaluated explicitly given $(s_{a,b}^{(z,w)})_{a,b,z,w}$ using Eqs.~\eqref{robust1stterm}--\eqref{robust3rdterm}, and that is (for small $\theta$) of order $\mathcal{O}(\theta)$---with $\mathcal{J}_{\text{correction}}=0$ when $\theta=0$.
Given the witness inequality $\mathcal{J}_{\text{ref.}} \ge 0$, which implies
\begin{align}
    \mathcal{J}_{\text{exp.}} \ge - \mathcal{J}_{\text{correction}}^{\text{max}}, \label{eq:robust_witness}
\end{align}
we thus find that a violation of Eq.~\eqref{eq:robust_witness} certifies causal nonseparability in a manner that is robust to noise.


\section{``Network-MDCI'' certification of all causally nonseparable process matrices}
\label{app:MDCI}

\renewcommand{\theequation}{D\arabic{equation}}
\setcounter{equation}{0}

\begin{figure}[t]
	\begin{center}
	\includegraphics[width=0.7\columnwidth]{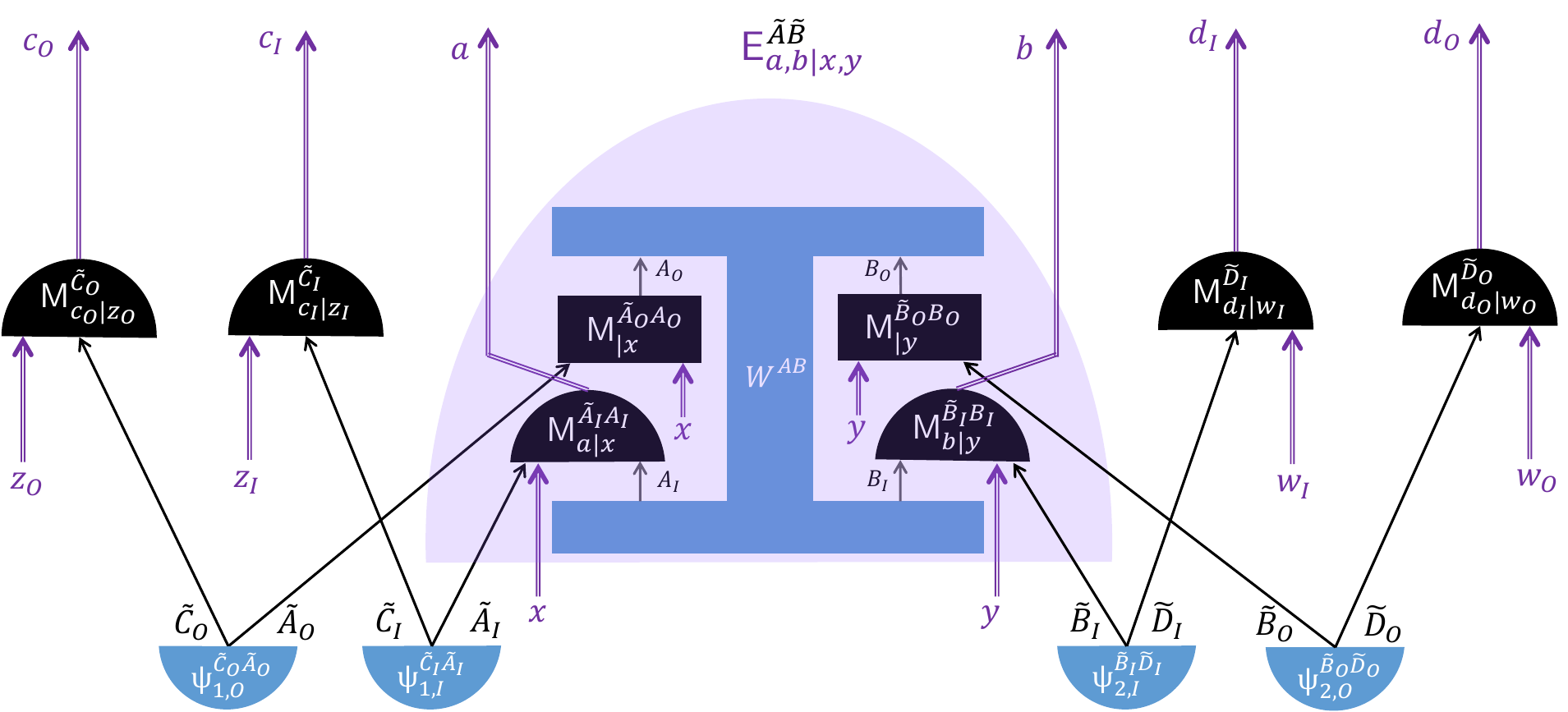}
	\end{center}
	\caption{Bipartite Network-MDCI scenario: Alice and Bob, instead of each having a single black-box instrument (cf.\ Fig.~\ref{fig:scenarii}, bottom), now each have two black-box devices (a channel and a measurement device, $\mathsf{M}_{a|x}^{\pAI A_I}$ and $\mathsf{M}_{|x}^{\pAO A_O}$ for Alice, and $\mathsf{M}_{b|y}^{\pBI B_I}$ and $\mathsf{M}_{|y}^{\pBO B_O}$ for Bob) that they plug into the process matrix $W^{AB}$ with the network connectivity shown here. The quantum inputs remotely prepared for these devices are independently self-tested in the same way as for the standard NDI scenario, which then allows one to certify the causal nonseparability of the D-POVM $(\mathsf{E}_{a,b|x,y}^{\pA\pB})_{a,b}$ (purple semicircle) for $x=y=\star$, and thus the causal nonseparability of the process matrix $W^{AB}$ in a N-MDCI scenario.}
	\label{fig:mdci_scenario}
\end{figure}

It was shown in~\cite{dourdent21} that all bipartite causally nonseparable process matrices can be certified with trusted quantum inputs, if we assume that Alice and Bob's (trusted) ancillary Hilbert spaces have a bipartite structure of the form $\HS^{\rA} = \HS^{\rAI\rAO}$ and $\HS^{\rB} = \HS^{\rBI\rBO}$, and that their instruments have a specific structure as well---namely, that they factorise as a joint measurement of system $\rAI$ together with the system $A_I$ received from the process matrix, and a channel taking system $\rAO$ into the system $A_O$ that is fed into the process matrix (and similarly on Bob's side).
For this certification, it suffices to use separate (product) quantum inputs in systems $\rAI$ and $\rAO$ (and in systems $\rBI$ and $\rBO$).
This certification, called the measurement-device-and-channel-independent (MDCI) certification, can be transformed into an NDI certification by making similar assumptions on the (now untrusted) physical spaces and instruments\changed{; see Fig.~\ref{fig:mdci_scenario}.}

Let us first recall that the above assumptions imply that the Choi matrices of Alice's instruments decompose as $\mathsf{M}_{a|x}^{\pA A} = \mathsf{M}_{a|x}^{\pAI A_I} \otimes \mathsf{M}_{|x}^{\pAO A_O}$, for some POVM $(\mathsf{M}_{a|x}^{\pAI A_I})_a$ and a channel $\mathsf{M}_{|x}^{\pAO A_O}$, and similarly for Bob. Given that the channels $\mathsf{M}_{|x}^{\pAO A_O}$ and $\mathsf{M}_{|y}^{\pBO B_O}$ must be trace-preserving, these decompositions imply that
\begin{align}
    \Tr_{A_O} \mathsf{M}_{a|x}^{\pA A} = \mathsf{M}_{a|x}^{\pAI A_I} \otimes \id^{\pAO}, \quad \Tr_{B_O} \mathsf{M}_{b|y}^{\pB B} = \mathsf{M}_{b|y}^{\pBI B_I} \otimes \id^{\pBO}, \label{eq:MDCI_assumption}
\end{align}
for all $a,b$, with $\sum_a \mathsf{M}_{a|x}^{\pAI A_I} = \id^{\pAI A_I}$ and $\sum_b \mathsf{M}_{b|y}^{\pBI B_I} = \id^{\pBI B_I}$. To obtain our results below, it is in fact sufficient to assume that Eq.~\eqref{eq:MDCI_assumption} holds, for some particular values of $a$ and $b$.

Consider a causally ordered process matrix $W = W^{A\prec B_I}\otimes \id^{B_O}$. The structure of Eq.~\eqref{eq:MDCI_assumption} implies that the physical D-POVM elements $\mathsf{E}_{a,b|x,y}^{\pA\pB}=\big(\mathsf{M}_{a|x}^{\pA A}\otimes \mathsf{M}_{b|y}^{\pB B}\big)*W^{AB}$ 
satisfy 
\begin{align}
    \mathsf{E}_{a,b|x,y}^{\pA\pB}=\big(\mathsf{M}_{a|x}^{\pA A}\otimes \Tr_{B_O} \mathsf{M}_{b|y}^{\pB B}\big)*W^{A\prec B_I} = \big(\mathsf{M}_{a|x}^{\pA A}\otimes \mathsf{M}_{b|y}^{\pBI B_I}\big)*W^{A\prec B_I} \otimes \id^{\pBO} = \mathsf{E}_{a,b|x,y}^{\pA\pBI}\otimes \id^{\pBO}
\end{align}
with $\mathsf{E}_{a,b|x,y}^{\pA\pBI} = \big(\mathsf{M}_{a|x}^{\pA A}\otimes \mathsf{M}_{b|y}^{\pBI B_I}\big)*W^{A\prec B_I}$.

Moving on to the NDI scenario, let us focus, for now, on the MDCI assumption on Bob's side. Instead of introducing just one source shared between Bob and Daisy, we now introduce two independent sources shared between Bob and two new parties, $\text{Daisy}_I$, with Hilbert space $\HS^{\pDI}$, and $\text{Daisy}_O$, with Hilbert space $\HS^{\pDO}$ (we use the same notations as before, just adding the subscripts ${}_I$ and ${}_O$ appropriately).
Self-testing the quantum inputs in spaces $\HS^{\pBI}$ and $\HS^{\pBO}$ separately, we can write, similarly to Eq.~\eqref{eq:MD_rhoBD},
\begin{align}
    P(d_I|w_I)\, \uprho_{d_I|w_I}^\pBI & = \pM_{d_I|w_I}^\pDI * \uppsi_{2,I}^{\pBI\pDI} = P(d_I|w_I)\, {V^{\mathbf{B}_I}}^\dagger\left( \sum_{l=0,1} (\rho_{d_I|w_I}^\rBI)^{T^{l}}\otimes\zeta_{l}^{\jfB_I} \right) V^{\mathbf{B}_I}, \notag \\
    P(d_O|w_O)\, \uprho_{d_O|w_O}^\pBO & = \pM_{d_O|w_O}^\pDO * \uppsi_{2,O}^{\pBO\pDO} = P(d_O|w_O)\, {V^{\mathbf{B}_O}}^\dagger\left( \sum_{l'=0,1} (\rho_{d_O|w_O}^\rBO)^{T^{l'}}\otimes\zeta_{l'}^{\jfB_O} \right) V^{\mathbf{B}_O},
\end{align}
for some reference quantum input states $\rho_{d_I|w_I}^\rBI \in \L(\HS^\rBI)$ and $\rho_{d_O|w_O}^\rBO \in \L(\HS^\rBO)$.

Then as in Eq.~\eqref{eq:P_cabd_zxyw}, and still for a causally ordered $W = W^{A\prec B_I}\otimes \id^{B_O}$, the correlations $P(c,a,b,d_I,d_O|z,x,y,w_I,w_O)$ can then be written (using the above equations together with Eq.~\eqref{eq:MC_rhoCA}) as
\begin{align}
& P(c,a,b,d_I,d_O|z,x,y,w_I,w_O) = \left(\!\pM_{c|z}^\pC \!\otimes\! \pM_{a|x}^{\pA A}\!\otimes\! \pM_{b|y}^{\pB B} \!\otimes\! \pM_{d_I|w_I}^\pDI \!\otimes\! \pM_{d_O|w_O}^\pDO\!\right) \!*\! \left(\!\uppsi_1^{\pC\pA}\!\otimes\! W^{A\prec B_I}\!\otimes\! \id^{B_O} \!\otimes\! \uppsi_{2,I}^{\pBI\pDI} \!\otimes\! \uppsi_{2,O}^{\pBO\pDO}\!\right) \notag \\[1mm]
& \quad = P(c|z)\, P(d_I|w_I) \, P(d_O|w_O) \, \left( \mathsf{E}_{a,b|x,y}^{\pA\pBI} \otimes \id^{\pBO} \right) * \Bigg(\sum_{k=0,1} \Big( \dketbra{{V^{\mathbf{A}}}^\dagger}{{V^{\mathbf{A}}}^\dagger}^{\rA\jfA\pA} * \xi_k^{\jfA} \Big)^{T_{\rA}^k} * \rho_{c|z}^\rA \notag \\[-2mm]
& \hspace{85mm} \otimes \sum_{l=0,1} \Big( \dketbra{{V^{\mathbf{B}_I}}^\dagger}{{V^{\mathbf{B}_I}}^\dagger}^{\rBI\jfB_I\pBI} * \zeta_{l}^{\jfB_I} \Big)^{T_{\rBI}^{l}} * \rho_{d_I|w_I}^\rBI \notag \\[-2mm]
& \hspace{90mm} \otimes \sum_{l'=0,1} \Big( \dketbra{{V^{\mathbf{B}_O}}^\dagger}{{V^{\mathbf{B}_O}}^\dagger}^{\rBO\jfB_O\pBO} * \zeta_{l'}^{\jfB_O} \Big)^{T_{\rBO}^{l'}} * \rho_{d_O|w_O}^\rBO \Bigg) \notag \\[1mm]
& \quad = P(c|z)\, P(d_I|w_I) \, P(d_O|w_O)\ \left( E_{a,b|x,y}^{\rA\rBI} \otimes \id^\rBO \right) * \left( \rho_{c|z}^\rA \otimes \rho_{d_I|w_I}^\rBI \otimes \rho_{d_O|w_O}^\rBO \right) \label{eq:P_cabdd_zxyww}
\end{align}
with
\begin{align}
    E_{a,b|x,y}^{\rA\rBI} & = \mathsf{E}_{a,b|x,y}^{\pA\pBI} * \Bigg[ \sum_{k=0,1} \left( \dketbra{{V^{\mathbf{A}}}^\dagger}{{V^{\mathbf{A}}}^\dagger}^{\rA\jfA\pA} * \xi_k^{\jfA} \right)^{T_{\rA}^k} \otimes \sum_{l=0,1} \left(\dketbra{{V^{\mathbf{B}_I}}^\dagger}{{V^{\mathbf{B}_I}}^\dagger}^{\rBI\jfB_I\pBI} * \zeta_{l}^{\jfB_I} \right)^{T_{\rBI}^{l}} \Bigg],
\end{align}
where we used the fact that $\sum_{l'} \Big( \Tr_{\pBO} \dketbra{{V^{\mathbf{B}_O}}^\dagger}{{V^{\mathbf{B}_O}}^\dagger}^{\rBO\jfB_O\pBO} * \zeta_{l'}^{\jfB_O} \Big)^{T_{\rBO}^{l'}} = \id^{\rBO}$, cf.\ Footnote~\ref{ftn:sum_trB_V_V_zeta}.

Hence, we see that for a causally ordered $W = W^{A\prec B_I}\otimes \id^{B_O}$, the structure $\mathsf{E}_{a,b|x,y}^{\pA\pB} = \mathsf{E}_{a,b|x,y}^{\pA\pBI}\otimes \id^{\pBO}$ that the MDCI assumption (on Bob's side) implies on the physical D-POVM elements $\mathsf{E}_{a,b|x,y}^{\pA\pB}$ gets transferred to the effective D-POVM elements $E_{a,b|x,y}^{\rA\rB} = E_{a,b|x,y}^{\rA\rBI} \otimes \id^\rBO$ that act on the reference input states. By symmetry, the same structure preservation applies for a causally ordered $W = W^{B\prec A_I}\otimes \id^{A_O}$ (using the MDCI assumption on Alice's side, introducing now two independent sources shared by two new parties $\text{Charlie}_I$ and $\text{Charlie}_O$).
It follows that for a causally separable process matrix $W^{AB} = q\,W^{A\prec B_I} \otimes \id^{B_O} + (1{-}q)\,W^{B\prec A_I} \otimes \id^{A_O}$, the D-POVM elements acting on the reference input states (in some reference Hilbert spaces $\HS^{\rAI}$, $\HS^{\rAO}$, $\HS^{\rBI}$ and $\HS^{\rBO}$) decompose as
\begin{align}
    E_{a,b|x,y}^{\rAI\rAO\rBI\rBO} & = q\, E_{a,b|x,y}^{\rAI\rAO\rBI} \otimes \id^\rBO + (1-q)\, E_{a,b|x,y}^{\rAI\rBI\rBO} \otimes \id^\rAO. \label{eq:N-MDCI}
\end{align}
In this scenario (involving all four additional parties $\text{Charlie}_I$, $\text{Charlie}_O$, $\text{Daisy}_I$ and $\text{Daisy}_O$ together with Alice and Bob), the correlations
\begin{align}
    P_\star(a,b|z,w) & = P(a,b|z_I,z_O, x{=}\star, y{=}\star, w_I,w_O, c_I{=}0, c_O{=}0, d_I{=}0, d_O{=}0) \notag \\
    & = E_{a,b|\star,\star}^{\rAI\rAO\rBI\rBO} * \left( \rho_{z_I}^\rAI \otimes \rho_{z_O}^\rAO \otimes \rho_{w_I}^\rBI \otimes \rho_{w_O}^\rBO \right)
\end{align}
then define a distribution $P'(a,b|\rho_{z_I}^\rAI, \rho_{z_O}^\rAO, \rho_{w_I}^\rBI, \rho_{w_O}^\rBO)$ in a SDI-QI setting with the reference quantum inputs, that must respect any witness inequality for correlations induced by D-POVM elements of the form of Eq.~\eqref{eq:N-MDCI}. Now, we have shown in~\cite{dourdent21} that any bipartite causally nonseparable process matrix can violate such a witness inequality. We thus conclude here that any bipartite causally nonseparable process matrix can be certified in the scenario considered here, which combines the NDI approach with the MDCI structure assumption---i.e., in a ``N-MDCI'' (``Network-MDCI'') manner.

Recall that in~\cite{dourdent21} we also studied MDCI certification in the ``($2$+$F$)-partite'' scenario (which involves an additional party, Fiona, in the causal future of Alice and Bob and who has no output Hilbert space), showing that a similar result holds for all ``TTU-'' and ``TUU-noncausal'' processes as defined in~\cite{bavaresco19}---that is, causally nonseparable processes that can be certified with trusted operations for Alice and Bob (but not for Fiona), or for Alice only.
In particular, we showed that all ``TTU-'' and ``TUU-noncausal'' processes could be certified in a ``MDCI-MDCI-DI'' and ``MDCI-DI-DI'' manner, respectively, when imposing the extra assumption of Eq.~\eqref{eq:MDCI_assumption} to the parties with formally trusted operations (and still with trusted quantum inputs). Following the same arguments as above, we can conclude that all such processes---including the quantum switch--can also be certified in a N-MDCI manner for those parties, thus relaxing the need for trusted quantum inputs.

Although we leave it as an open problem, we expect it to be rather straightforward to further generalise all these results to the ($P$+$2$+$F$)-partite scenario, which also includes a party in the global past and that we will consider in Appendix~\ref{app:sdiqiP2F}.


\section{SDI-QI and NDI certification of causal nonseparability in the ``($P$+$2$+$F$)-partite'' scenario}
\label{app:sdiqiP2F}

\renewcommand{\theequation}{E\arabic{equation}}
\setcounter{equation}{0}

We consider here the ``($P$+$2$+$F$)-partite'' scenario introduced in the main text, which extends the ($2$+$F$)-partite scenario (considered in particular in~\cite{dourdent21}) by adding a new party, Phil, in the causal past of all other parties. Phil is only connected to the process matrix through his output Hilbert space $\HS^{P}$ (he has no input Hilbert space). We shall first recall the definition of causal nonseparability in this scenario, before considering its SDI-QI certification when all four parties receive quantum inputs, as well as when only Phil does, and applying this to the quantum switch.
As in the bipartite scenario, this SDI-QI certification can be readily transformed into NDI certification, and we will briefly discuss this transformation, which is performed analogously to the bipartite case, at the end of the section.


\subsection{Causal (non)separability in the ``($P$+$2$+$F$)-partite'' scenario}
\label{app:cnsp1}

Given that Phil is in the global past of the process, and that Fiona is in the global future, the only relevant orders in this scenario are $P\prec A\prec B\prec F$ and $P\prec B\prec A\prec F$.
The causally separable process matrices are those that can be decomposed as~\cite{wechs19}
\begin{align}
W^{PABF} & = W^{P\prec A\prec B\prec F} + W^{P\prec B\prec A\prec F} \label{eq:csep_PABF}
\end{align}
such that
\begin{align}
    \Tr_F W^{P\prec A\prec B\prec F} = W^{P\prec A\prec B_I} \otimes \id^{B_O}, & \qquad \Tr_F W^{P\prec B\prec A\prec F} = W^{P\prec B\prec A_I} \otimes \id^{A_O}, \notag \\
    \Tr_{B_I} W^{P\prec A\prec B_I} = W^{P\prec A_I} \otimes \id^{A_O}, & \qquad \Tr_{A_I} W^{P\prec B\prec A_I} = W^{P\prec B_I} \otimes \id^{B_O}, \notag \\
    \Tr_{A_I} W^{P\prec A_I} + &\Tr_{B_I}W^{P\prec B_I} = \id^{P}, \label{eq:csep_PABF_decomp}
\end{align}
with positive semidefinite matrices $W^{P\prec A\prec B_I}\in\L(\HS^{PAB_I})$, $W^{P\prec B\prec A_I}\in\L(\HS^{PBA_I})$, $W^{P\prec A_I}\in\L(\HS^{PA_I})$, $W^{P\prec B_I}\in\L(\HS^{PB_I})$.

Note that in general $W^{P\prec A\prec B\prec F}$ and $W^{P\prec B\prec A\prec F}$ are not necessarily valid (subnormalised) process matrices themselves (unless $\Tr_{A_I} W^{P\prec A_I}\propto \id^P$ and $\Tr_{B_I}W^{P\prec B_I}\propto\id^P$). Indeed causally separable processes allow for the order between Alice and Bob to be influenced by Phil's action (in a classically controlled manner~\cite{wechs21}), rather than involving only convex combinations of processes with the fixed orders $P\prec A\prec B\prec F$ and $P\prec B\prec A\prec F$.
Such causally separable processes with dynamical causal order thus cannot be written as a convex mixture of valid processes with fixed causal orders. This is why, contrarily to Eqs.~\eqref{eq:csep2} and~\eqref{eq:N-MDCI}, we did not include convex weights in Eq.~\eqref{eq:csep_PABF}, which would otherwise suggest such a combination.


\subsection{SDI-QI certification with quantum inputs for all 4 parties}
\label{app:cnsp2}

\begin{figure}[t!]
	\begin{center}
	\includegraphics[width=0.4\columnwidth]{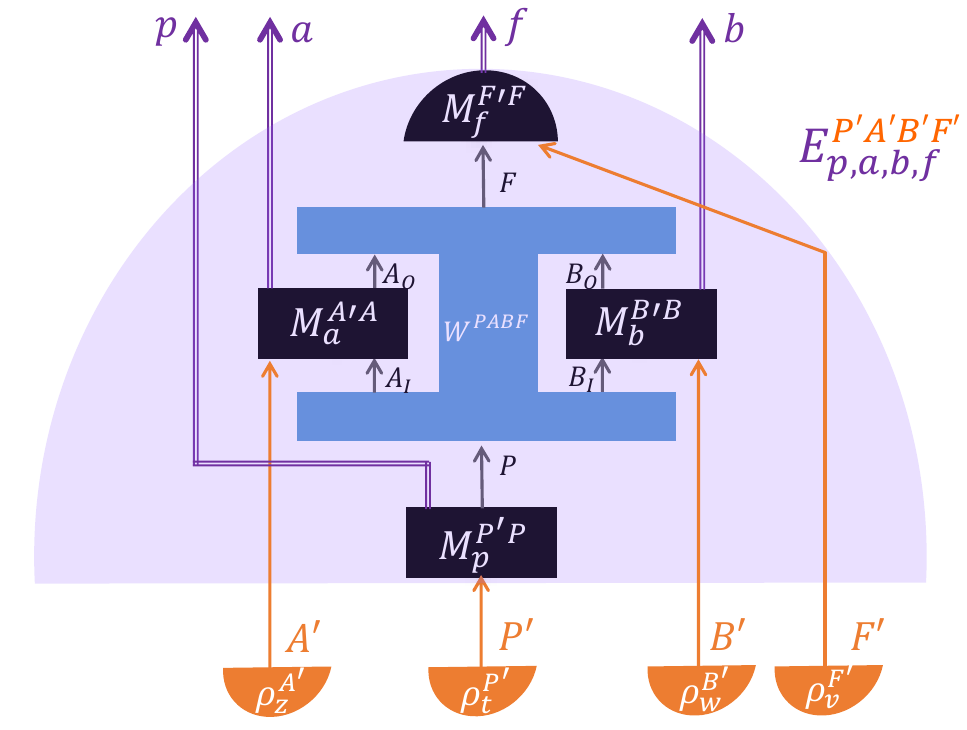}
	\end{center}
	\caption{($P$+$2$+$F$)-partite SDI-QI scenario with quantum inputs for Phil, Alice, Bob and Fiona. The trusted quantum inputs (orange) $\rho_t^{\rP}\in\L(\HS^{\rP}),\rho_z^{\rA}\in\L(\HS^{\rA}),\rho_w^{\rB}\in\L(\HS^{\rB}),\rho_v^{\rF}\in\L(\HS^{\rF})$ are prepared and effectively measured by the D-POVM $(E^{\rP\rA\rB\rF}_{p,a,b,f})_{p,a,b,f}$ generated by Phil, Alice, Bob and Fiona's untrusted operations on the process matrix $W^{PABF}$.}
\label{fig:sdiqipf}
\end{figure}

In a general ($P$+$2$+$F$)-partite SDI-QI scenario, all four parties Phil, Alice, Bob and Fiona receive some quantum inputs $\rho_t^{\rP}$, $\rho_z^{\rA}$, $\rho_w^{\rB}$ and $\rho_v^{\rF}$ in some Hilbert spaces $\L(\HS^\rP)$, $\L(\HS^\rA)$, $\L(\HS^\rB)$ and $\L(\HS^\rF)$ respectively; see Fig.~\ref{fig:sdiqipf}. Generalising the bipartite case introduced in the main text, the correlations established by the four parties are
\begin{align}
P(p,a,b,f|\rho_t^\rP,\rho_z^\rA,\rho_w^\rB,\rho_v^\rF) & = \big(M_p^{\rP P}\otimes M_a^{\rA A}\otimes M_b^{\rB B}\otimes M_f^{\rF F}\big) * \big(\rho_t^\rP\otimes\rho_z^\rA\otimes\rho_w^\rB\otimes\rho_v^\rF\otimes W^{PABF}\big) \notag \\
 & = E_{p,a,b,f}^{\rP\rA\rB\rF} * \big(\rho_t^\rP\otimes\rho_z^\rA\otimes\rho_w^\rB\otimes\rho_v^\rF\big) ,
\end{align}
where we introduced the D-POVM elements
\begin{align}
E_{p,a,b,f}^{\rP\rA\rB\rF} = \big(M_p^{\rP P}\otimes M_a^{\rA A}\otimes M_b^{\rB B}\otimes M_f^{\rF F}\big)*W^{PABF}.
\end{align}

Assuming that $W^{PABF}$ is a causally separable process matrix satisfying Eqs.~\eqref{eq:csep_PABF}--\eqref{eq:csep_PABF_decomp}, the induced D-POVM $\mathbb{E}^{\rP\rA\rB\rF}=(E_{p,a,b,f}^{\rP\rA\rB\rF})_{p,a,b,f}$ also decomposes as
\begin{align}
    \mathbb{E}^{\rP\rA\rB\rF} = \mathbb{E}^{\rP\prec\rA\prec\rB\prec\rF} + \mathbb{E}^{\rP\prec\rB\prec\rA\prec\rF}
    \label{eq:Epabf_csep1}
\end{align}
with $\mathbb{E}^{\rP\prec\rA\prec\rB\prec\rF}=(E_{p,a,b,f}^{\rP\prec\rA\prec\rB\prec\rF})_{p,a,b,f}$ and $\mathbb{E}^{\rP\prec\rB\prec\rA\prec\rF}=(E_{p,a,b,f}^{\rP\prec\rB\prec\rA\prec\rF})_{p,a,b,f}$ such that
\begin{align}
    E_{p,a,b,f}^{\rP\prec\rA\prec\rB\prec\rF} & = (M_p^{\rP P}\otimes M_a^{\rA A}\otimes M_b^{\rB B}\otimes M_f^{\rF F})*W^{P\prec A\prec B\prec F}, \notag \\[2mm]
    \sum_f E_{p,a,b,f}^{\rP\prec\rA\prec\rB\prec\rF} & = (M_p^{\rP P}\otimes M_a^{\rA A}\otimes \Tr_{B_O}M_b^{\rB B})*W^{P\prec A\prec B_I} \otimes \id^{\rF} = E_{p,a,b}^{\rP\prec\rA\prec\rB} \otimes \id^{\rF}, \notag \\
    \sum_b E_{p,a,b}^{\rP\prec\rA\prec\rB} & = (M_p^{\rP P}\otimes \Tr_{A_O}M_a^{\rA A})*W^{P\prec A_I} \otimes \id^{\rB} = E_{p,a}^{\rP\prec\rA} \otimes \id^{\rB}, \notag \\
    \sum_a E_{p,a}^{\rP\prec\rA} & = M_p^{\rP P}*\Tr_{A_I}W^{P\prec A_I} \otimes \id^{\rA} = E_{p}^{\rP\,[A\prec B]} \otimes \id^{\rA}, \label{eq:Epabf_csep2}
\end{align}
analogously for $E_{p,a,b,f}^{\rP\prec\rB\prec\rA\prec\rF}$ and  the order $P\prec B\prec A\prec F$, and with 
\begin{align}
    E_{p}^{\rP\,[A\prec B]} + E_{p}^{\rP\,[B\prec A]} & = M_p^{\rP P}*\id^P = \Tr_P M_p^{\rP P}, \label{eq:Epabf_csep3}
\end{align}
such that, since $\sum_p M_p^{\rP P}$ is a CPTP map,
\begin{align}
    \sum_p (E_{p}^{\rP\,[A\prec B]} + E_{p}^{\rP\,[B\prec A]}) & = \id^{\rP}. \label{eq:Epabf_csep4}
\end{align}
Note that the individual sets of positive semidefinite matrices $\mathbb{E}^{\rP\prec\rA\prec\rB\prec\rF}=(E_{p,a,b,f}^{\rP\prec\rA\prec\rB\prec\rF})_{p,a,b,f}$, $\mathbb{E}^{\rP\prec\rB\prec\rA\prec\rF}=(E_{p,a,b,f}^{\rP\prec\rB\prec\rA\prec\rF})_{p,a,b,f}$, $\mathbb{E}^{\rP\prec\rA\prec\rB}=(E_{p,a,b}^{\rP\prec\rA\prec\rB})_{p,a,b}$, $\mathbb{E}^{\rP\prec\rB\prec\rA}=(E_{p,a,b}^{\rP\prec\rB\prec\rA})_{p,a,b}$, $\mathbb{E}^{\rP\prec\rA}=(E_{p,a}^{\rP\prec\rA})_{p,a}$, $\mathbb{E}^{\rP\prec\rB}=(E_{p,b}^{\rP\prec\rB})_{p,b}$, $\mathbb{E}^{\rP\,[A\prec B]}=(E_{p}^{\rP\,[A\prec B]})_{p}$, $\mathbb{E}^{\rP\,[B\prec A]}=(E_{p}^{\rP\,[B\prec A]})_{p}$ are not D-POVMs as they do not sum to the identity in general. On the other hand, the set $\mathbb{E}^{\rP\,[A\prec B]}\cup\mathbb{E}^{\rP\,[B\prec A]}$ formed by all elements of $\mathbb{E}^{\rP\,[A\prec B]}$ and $\mathbb{E}^{\rP\,[B\prec A]}$ taken together forms a valid POVM.

Analogously to the bipartite and ($2$+$F$)-partite cases, this leads us to the following definition.
\begin{definition}\label{def:csep_DPOVM_P2F}
A ($P$+$2$+$F$)-partite D-POVM $\mathbb{E}^{\rP\rA\rB\rF} = (E_{p,a,b,f}^{\rP\rA\rB\rF})_{p,a,b,f}$ that can be decomposed in the form
\begin{align}
\mathbb{E}^{\rP\rA\rB\rF} = \mathbb{E}^{\rP\prec\rA\prec\rB\prec\rF} + \mathbb{E}^{\rP\prec\rB\prec\rA\prec\rF},
\end{align}
where $\mathbb{E}^{\rP\prec\rA\prec\rB\prec\rF} = (E_{p,a,b,f}^{\rP\prec\rA\prec\rB\prec\rF})_{p,a,b,f}$ and $\mathbb{E}^{\rP\prec\rB\prec\rA\prec\rF} = (E_{p,a,b,f}^{\rP\prec\rB\prec\rA\prec\rF})_{p,a,b,f}$ satisfy
\begin{align}
    \sum_f E_{p,a,b,f}^{\rP\prec\rA\prec\rB\prec\rF} = E_{p,a,b}^{\rP\prec\rA\prec\rB} \otimes \id^{\rF}, \quad &
    \sum_b E_{p,a,b}^{\rP\prec\rA\prec\rB} = E_{p,a}^{\rP\prec\rA} \otimes \id^{\rB}, \quad
    \sum_a E_{p,a}^{\rP\prec\rA} = E_{p}^{\rP\,[A\prec B]} \otimes \id^{\rA}, \notag \\
    \sum_f E_{p,a,b,f}^{\rP\prec\rB\prec\rA\prec\rF} = E_{p,a,b}^{\rP\prec\rB\prec\rA} \otimes \id^{\rF}, \quad &
    \sum_a E_{p,a,b}^{\rP\prec\rB\prec\rA} = E_{p,b}^{\rP\prec\rB} \otimes \id^{\rA}, \quad
    \sum_b E_{p,b}^{\rP\prec\rB} = E_{p}^{\rP\,[B\prec A]} \otimes \id^{\rB}, \notag \\
    & \sum_p (E_{p}^{\rP\,[A\prec B]} + E_{p}^{\rP\,[B\prec A]}) = \id^{\rP}
\end{align}
(with all operators $E_{\cdots}^{\cdots} \ge 0$) is said to be \emph{causally separable}.
\end{definition}

As noted above, a causally separable ($P$+$2$+$F$)-partite process matrix can only generate a causally separable ($P$+$2$+$F$)-partite D-POVM. \emph{A contrario}, if one finds that, for some choice of operations for each party, the induced D-POVM is causally nonseparable, this certifies in a SDI-QI manner that the process matrix that generated it is itself causally nonseparable. Similarly to the bipartite and the ($2$+$F$)-partite cases, one can construct a causal witness---and associated witness inequality---that can be measured by providing tomographically complete quantum inputs to the parties, thereby allowing one to certify the causal non-separability of ($P$+$2$+$F$)-partite D-POVMs~\cite{araujo15,branciard16a,dourdent21}.


\subsection{SDI-QI certification with quantum inputs only for Phil}
\label{app:cnsp3}

SDI-QI scenarios can also be considered, where not all parties receive quantum inputs: some may receive classical inputs (or no input at all). For example, in~\cite{dourdent21} we considered the ($2$+$F$)-partite case where only Alice and Bob (but not Fiona) receive quantum inputs---a situation that could readily be extended to the ($P$+$2$+$F$)-partite case.

\begin{figure}[t!]
	\begin{center}
	\includegraphics[width=0.45\columnwidth]{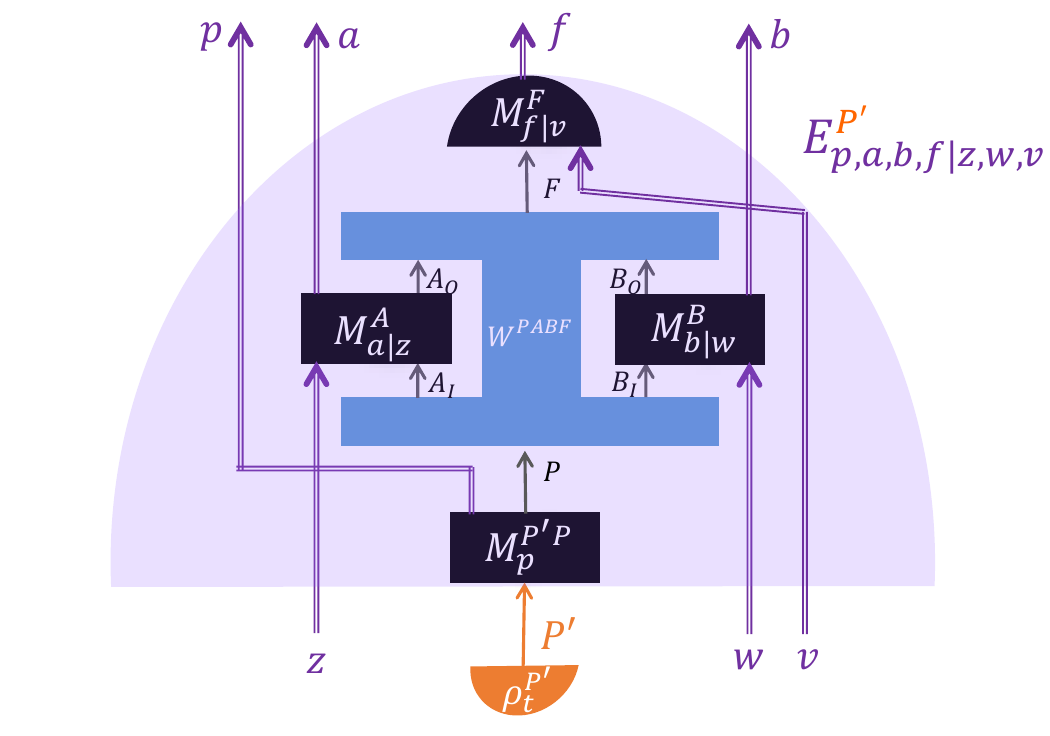}
	\end{center}
	\caption{($P$+$2$+$F$)-partite SDI-QI scenario with quantum inputs solely for Phil. The trusted quantum input (orange) $\rho_t\in\L(\HS^{\rP})$ is prepared and effectively measured by the D-POVMs $(E^{\rP}_{p,a,b,f|z,w,v})_{p,a,b,f}$ generated by Phil, Alice, Bob and Fiona's untrusted operations on the process matrix $W^{PABF}$.}
\label{fig:sdiqip}
\end{figure}

Let us here consider instead the case where only Phil receives quantum input states $\rho_t^{\rP}\in\L(\HS^{\rP})$, while Alice, Bob and Fiona are given classical inputs $z$, $w$ and $v$ respectively; see Fig.~\ref{fig:sdiqip}.
Conditioning their instruments explicitly on the classical inputs,%
\footnote{Classical inputs can also be fitted into the framework with quantum inputs by using orthogonal quantum input states. Indeed, one can encode (e.g., for Alice) the classical inputs into some orthonormal basis $\{\ket{z}^\rA\}_z$, and introduce the instruments $(M_a^{\rA A})_a = \big( \sum_z \ketbra{z}{z}^\rA\otimes M_{a|z}^A\big)_a$~\cite{dourdent21}. One can use this, for instance, to obtain the analogous constraints to Eqs.~\eqref{eq:Epabf_csep1}--\eqref{eq:Epabf_csep4} for classical inputs $z,w,v$, and to thereby justify the decomposition in Definition~\ref{def:csep_DPOVM_P2F_zwv}---and in particular, the ``no-signalling constraints'' (e.g., that $\sum_f E_{p,a,b,f|z,w,v}^{\rP\,[A\prec B]} = E_{p,a,b|z,w}^{\rP\,[A\prec B]}$ does not depend on $v$, that $\sum_b E_{p,a,b|z,w}^{\rP\,[A\prec B]} = E_{p,a|z}^{\rP\,[A\prec B]}$ does not depend on $w$, etc.).}
the correlations established by the four parties are given here by
\begin{align}
P(p,a,b,f|\rho_t^\rP,z,w,v) & = \big(M_p^{\rP P}\otimes M_{a|z}^A\otimes M_{b|w}^B\otimes M_{f|v}^F\big) * \big(\rho_t^\rP \otimes W^{PABF}\big) \notag \\
 & = E_{p,a,b,f|z,w,v}^{\rP} * \rho_t^\rP,
\end{align}
with the D-POVM elements%
\footnote{While the D-POVM acts on a single Hilbert space $\HS^\rP$, accessed by Phil only, it is still distributed in the sense that the classical inputs and outputs are associated with distributed parties, even if no quantum systems are associated with them.}
\begin{align}
E_{p,a,b,f|z,w,v}^{\rP} = \big(M_p^{\rP P}\otimes M_{a|z}^A\otimes M_{b|w}^B\otimes M_{f|v}^F\big)*W^{PABF}.
\end{align}

Assuming that $W^{PABF}$ is causally separable, one obtains a similar decomposition to Eqs.~\eqref{eq:Epabf_csep1}--\eqref{eq:Epabf_csep4}, for the whole family of D-POVMs $\{\mathbb{E}_{|z,w,v}^{\rP} = (E_{p,a,b,f|z,w,v}^{\rP})_{p,a,b,f}\}_{z,w,v}$ (i.e., involving the D-POVMs for all classical inputs $z,w,v$). Similarly to the previous case, we introduce here the following definition.
\begin{definition}\label{def:csep_DPOVM_P2F_zwv}
A family of ($P$+$2$+$F$)-partite D-POVMs $\{\mathbb{E}_{|z,w,v}^{\rP} = (E_{p,a,b,f|z,w,v}^{\rP})_{p,a,b,f}\}_{z,w,v}$ that can be decomposed in the form
\begin{align}
\mathbb{E}_{|z,w,v}^{\rP} = \mathbb{E}_{|z,w,v}^{\rP\,[A\prec B]} + \mathbb{E}_{|z,w,v}^{\rP\,[B\prec A]}, \label{eq:csep_DPOVM_P2F_zwv_1}
\end{align}
for all $z,w,v$, where $\mathbb{E}_{|z,w,v}^{\rP\,[A\prec B]} = (E_{p,a,b,f|z,w,v}^{\rP\,[A\prec B]})_{p,a,b,f}$ and $\mathbb{E}_{|z,w,v}^{\rP\,[B\prec A]} = (E_{p,a,b,f|z,w,v}^{\rP\,[B\prec A]})_{p,a,b,f}$ satisfy
\begin{align}
    \sum_f E_{p,a,b,f|z,w,v}^{\rP\,[A\prec B]} = E_{p,a,b|z,w}^{\rP\,[A\prec B]}, \quad &
    \sum_b E_{p,a,b|z,w}^{\rP\,[A\prec B]} = E_{p,a|z}^{\rP\,[A\prec B]}, \quad
    \sum_a E_{p,a|z}^{\rP\,[A\prec B]} = E_{p}^{\rP\,[A\prec B]}, \notag \\
    \sum_f E_{p,a,b,f|z,w,v}^{\rP\,[B\prec A]} = E_{p,a,b|z,w}^{\rP\,[B\prec A]}, \quad &
    \sum_a E_{p,a,b|z,w}^{\rP\,[B\prec A]} = E_{p,b|w}^{\rP\,[B\prec A]}, \quad
    \sum_b E_{p,b|w}^{\rP\,[B\prec A]} = E_{p}^{\rP\,[B\prec A]}, \notag \\
    & \sum_p (E_{p}^{\rP\,[A\prec B]} + E_{p}^{\rP\,[B\prec A]}) = \id^{\rP} \label{eq:csep_DPOVM_P2F_zwv_2}
\end{align}
(with all operators $E_{\cdots}^{\cdots} \ge 0$) is said to be \emph{causally separable}.
\end{definition}

As always, a causally separable ($P$+$2$+$F$)-partite process matrix $W^{PABF}$ can only generate a causally separable family of ($P$+$2$+$F$)-partite D-POVMs, so that obtaining a causally nonseparable family of ($P$+$2$+$F$)-partite D-POVMs (which can be verified using some form of causal witnesses) certifies in a SDI-QI manner the causal nonseparability of $W^{PABF}$.


\subsection{SDI-QI certification of the quantum switch in the ``($P$+$2$+$F$)-partite'' scenario}
\label{app:cnspqs}

The quantum switch~\cite{chiribella13} is the canonical example of a quantum process with indefinite causal order, which can be understood as a quantum circuit with quantum control of causal order~\cite{wechs21}. Despite being causally nonseparable, however, the quantum switch can only generate so-called causal correlations~\cite{araujo15,oreshkov16}, so that its causal nonseparability cannot be certified in the standard DI way.

As recalled previously, in~\cite{dourdent21} we showed that the causal nonseparability of the quantum switch could nevertheless be certified in a ($2$+$F$)-partite SDI-QI scenario where Alice and Bob receive quantum inputs.
This scenario can be seen as a particular case of a ($P$+$2$+$F$)-partite one, with a trivial role for the past party Phil. Introducing a nontrivial $P$ allows one to also provide the latter with quantum inputs. We will show here that the causal nonseparability of the quantum switch can still be certified if only Phil receives quantum inputs.

In the most general form of the quantum switch, as described in the main text, Phil sends and Fiona receives both ``target'' and ``control'' systems, in spaces $\HS^{P_\text{t}}$, $\HS^{P_\text{c}}$, $\HS^{F_\text{t}}$ and $\HS^{F_\text{c}}$, respectively.
Here, let us take the target system to be initialised in the state $\ketbra{0}{0}^{P_\text{t}}\in\L(\HS^{P_\text{t}})$ and traced out in the future space $\HS^{F_\text{t}}$, and the qubit control system to be prepared by Phil in the global past and given to Fiona in the global future.
The quantum switch can then be written as a ($P$+$2$+$F$)-partite process matrix as
\begin{equation}
W_\text{QS} = \ketbra{0}{0}^{P_t} * \Tr_{F_\text{t}} (\ketbra{w_\text{QS}}{w_\text{QS}}^{PP_\text{t}ABF_\text{t}F}),
\end{equation}
with
\begin{equation}
\ket{w_\text{QS}}^{PP_\text{t}ABF_\text{t}F} = \ket{0}^P\dket{\id}^{P_\text{t} A_I}\dket{\id}^{A_OB_I}\dket{\id}^{B_OF_\text{t}}\ket{0}^F  + \ket{1}^P\dket{\id}^{P_\text{t} B_I}\dket{\id}^{B_OA_I}\dket{\id}^{A_OF_\text{t}}\ket{1}^F
\label{eq:Ms_QSc}
\end{equation}
where $\dket{\id}^{XX'} = \sum_i \ket{i}^{X}\otimes\ket{i}^{X'}$ (and with implicit tensor products)~\cite{wechs21}.\footnote{When considering isomorphic Hilbert spaces $\HS^X$ and $\HS^{X'}$, we take their computational bases $\{\ket{i}^{X^{(\prime)}}\}_i$ to be in one-to-one correspondence.}

\begin{figure}[t]
	\begin{center}
	\includegraphics[width=0.4\columnwidth]{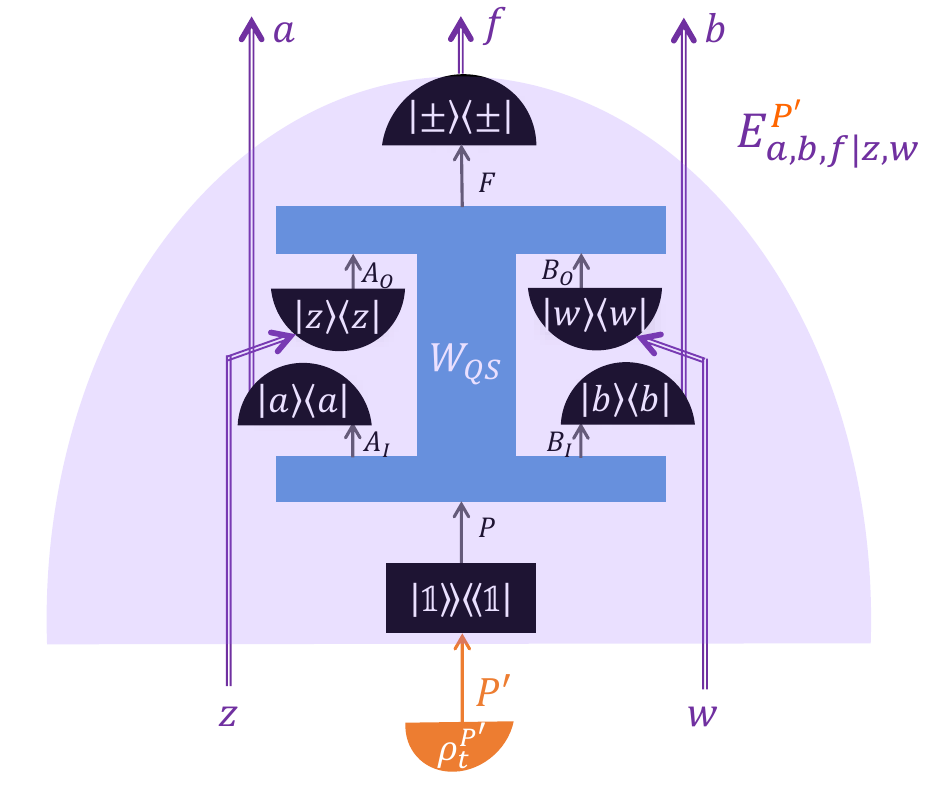}
	\end{center}
	\caption{D-POVM generated by the quantum instruments of Eq.~\eqref{eq:instruqs}, which can be used to certify the causal nonseparability of the quantum switch in a SDI-QI manner.}
\label{fig:sdiqipqs}
\end{figure}

Let us then consider the operations
\begin{align}
    M^{\rP P} = \dketbra{\id}{\id}^{\rP P}, \quad M_{a|z}^A = \ketbra{a}{a}^{A_I}\otimes\ketbra{z}{z}^{A_O}, \quad M_{b|w}^B = \ketbra{b}{b}^{B_I}\otimes\ketbra{w}{w}^{B_O}, \quad M_{f=\pm}^F = \ketbra{\pm}{\pm}^F,
    \label{eq:instruqs}
\end{align}
which can be interpreted as follows (see Fig.~\ref{fig:sdiqipqs}): 
 Phil simply implements an identity channel that directly forwards the quantum input into the quantum switch, with no (or just a trivial) output $p$. Alice and Bob measure the target system (taken here to be a qubit) in the computational basis with classical outcomes $a,b = 0,1$, and re-prepare it in the computational basis state that corresponds to their input $z,w = 0,1$. Fiona has no (or just a trivial) input $v$ and she simply measures the control qubit in the $\{\ket{\pm} = \frac{1}{\sqrt{2}}(\ket{0}\pm\ket{1})\}_\pm$ basis.

As it turns out, when implemented on a quantum switch, these instruments generate a causally nonseparable family of D-POVMs $\{\mathbb{E}_{|z,w}^{\rP} = (E_{a,b,f|z,w}^{\rP})_{a,b,f}\}_{z,w}$, according to Definition~\ref{def:csep_DPOVM_P2F_zwv} (with the trivial $p$ and $v$ omitted in this special case). Indeed, using a semidefinite programming approach, we can show that these D-POVMs cannot be decomposed as in Eqs.~\eqref{eq:csep_DPOVM_P2F_zwv_1}--\eqref{eq:csep_DPOVM_P2F_zwv_2}, and a causal witness can be constructed to test this~\cite{araujo15,branciard16a,dourdent21}.
More specifically, we find that a ``depolarised'' version of the quantum switch $W_\text{QS}(r) = {\textstyle \frac{1}{1+r}}(W_\text{QS} + r \, \id^\circ )$, obtained by mixing it with some fully white noise described by the process matrix $\id^\circ = \id^{PABF}/8$ (by some amount $r\ge 0$), induces with the above instruments a causally nonseparable family of D-POVMs for all $r< 2-\sqrt{2}$  (up to numerical precision).

Hence, as we claimed, the quantum switch can be certified in a SDI-QI manner by simply letting Phil receive quantum inputs. Of course, providing more parties with quantum inputs (or here, also providing Fiona with some input, or letting Phil also have outputs) could only make the certification more efficient.

Finally, as for any bipartite SDI-QI scenario considered in this work, the SDI-QI certification of a causally nonseparable process in the ($2$+$F$)- and ($P$+$2$+$F$)-partite scenarios can readily be turned into an NDI certification.
Indeed, by introducing new parties that share maximally entangled states with whichever parties receive quantum inputs, the quantum inputs can be independently self-tested as in Appendix~\ref{app:st} by self-testing these states and the new parties' measurements.
The proof then that the self-testing isometries associated with each quantum input preserve the causal separability of the D-POVM (i.e., that the effective D-POVM one obtains analogous to Eq.~\eqref{eq:rE_from_pE}) is analogous to that for the bipartite scenario given in Appendix~\ref{app:CNSep_NDI}.
For the quantum switch, in the case where only Phil receives a quantum input that he directly feeds into the quantum switch, as considered above (see Eq.~\eqref{eq:Ms_QSc}), this leads to a very similar scenario to that considered in Ref.~\cite{lugt22}, which we shall discuss in the next appendix.


\section{Comparing the NDI and DRF certifications of the quantum switch}
\label{app:DRF}

\renewcommand{\theequation}{F\arabic{equation}}
\setcounter{equation}{0}

In this appendix we provide some further comparison between our NDI certification of the causal nonseparability of the quantum switch and the DRF certification of Ref.~\cite{lugt22}. This complements the discussion in the main text of the underlying assumptions behind these two approaches.

\begin{figure}[ht!]
	\begin{center}
	\includegraphics[width=0.45\columnwidth]{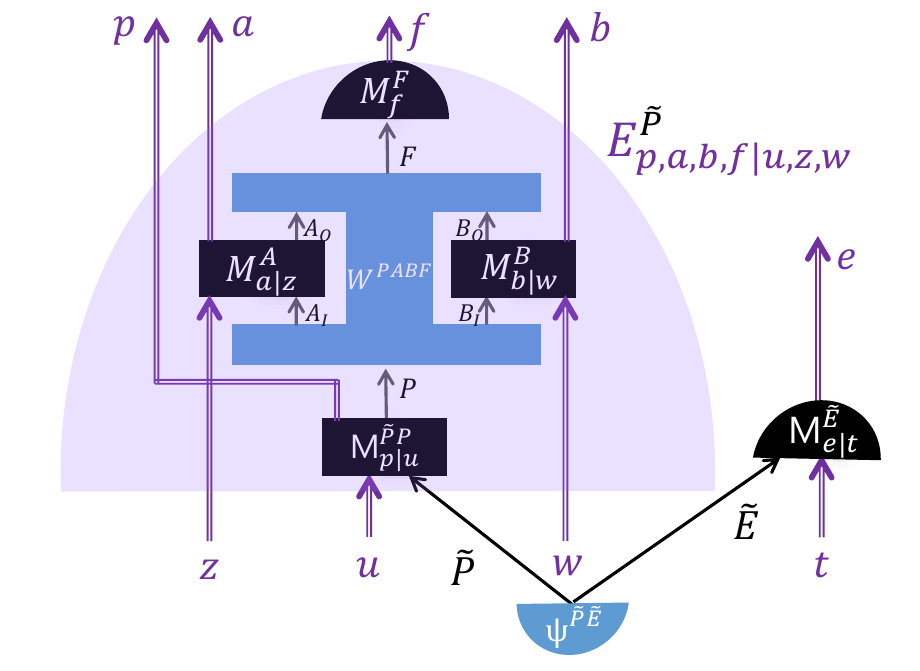}
	\end{center}
	\caption{Scenario for the NDI certification of the quantum switch with a global past. The remote preparation of a quantum input $\rho_t^{\rP}$ is self-tested by first self-testing the (untrusted) shared state $\uppsi^{\pP\tilde{E}}$ (to contain a maximally entangled state) and Emily's measurements $\pM_{e|t}^{\tilde{E}}$ (to effectively be tomographically complete projectors) from the correlations $P(p,e|u,t)$ shared by Phil and Emily.
	One can then certify the causal nonseparability of the family of D-POVMs $\{(E_{p,a,b,f|u,z,w}^{\tilde{P}})_{p,a,b,f}\}_{u,z,w}$ and thus the causal nonseparability of the process matrix $W^{PABF}$.}
	\label{fig:ndip}
\end{figure}

As mentioned at the end of Appendix~\ref{app:cnspqs}, the SDI-QI certification of the quantum switch described therein can be transformed into an NDI certification by following the same procedure as for the bipartite scenario discussed in detail in the main text.
In particular, the SDI-QI certification is transformed first into a NSDI-QI certification by adding an extra separated party, Emily, who shares a maximally entangled state with Phil and remotely prepares his quantum inputs $\rho_t^{\rP}$, and then into an NDI certification by self-testing both the state Phil and Emily share, and Emily's measurements.
The certification is then obtained by violating a witness inequality using the observed correlations $P_*(p,a,b,f|t,z,w) := P(p,a,b,f|u{=}*,t,z,w,e{=}0)=P'(p,a,b,f|\rho_t^{\rP},z,w)$, analogously to the bipartite setting.
This NDI scenario is shown in Fig.~\ref{fig:ndip}. 

As we already noted, Ref.~\cite{lugt22} showed how to device-independently certify the causal indefiniteness of the quantum switch by the violation of an inequality derived from the assumptions of Definite causal order, Relativistic causality, and Free interventions.
The structure of the ``DRF scenario'' they considered is very similar to that of our NDI certification, but there are some important differences.
In the DRF scenario, a (space-like) separated Emily is also added to a standard quantum switch scenario.
However, the scenario contains no Phil and, in contrast to the NDI certification, Fiona must be provided with a classical input (which we could also have done, but did not need to).
By having Alice share an entangled state with the control system in the global past of a quantum switch (directly incorporated in the process, instead of being provided by Phil), the parties can generate correlations violating the DRF inequality.

The DRF inequality itself, rephrased here in notation consistent with Fig.~\ref{fig:ndip} and with $v$ denoting Fiona's classical input, has the form~\cite{lugt22}
\begin{align}
    \mathcal{I}_\text{DRF}=P(e=0,b=z|t=0)+P(e=1,a=w|t=0)+P(f\oplus e=vt|z=w=0)\leq \frac{7}{4}.
    \label{eq:lcineq}
\end{align}
One can interpret this inequality as certifying (i) a space-like separated ``control of causal order'' (the first two terms of $\mathcal{I}_\text{DRF}$), and (ii) the non-classical, indeterministic behaviour of the control (the last term of $\mathcal{I}_\text{DRF}$). 
The latter of these can be interpreted as a CHSH inequality tested between Fiona (acting after the quantum switch, and who is thus provided with inputs) and Emily. 
It can be violated if the control system passes transparently through the process, which can be ensured to be the case if Alice and Bob perform appropriate operations (on inputs $z=w=0$).
On the other hand, while the NDI certification also relies on the violation of a Bell inequality, it is tested between Emily and Phil---who is in the causal past of the quantum switch and has classical inputs and outputs---to self-test the quantum state they share.
Fiona, in this case, needs no classical input and performs a fixed measurement.
Curiously, however, we were able to obtain NDI certification using exactly the same measure-and-prepare instruments for Alice and Bob (see Eq.~\eqref{eq:instruqs}) as were used to violate the DRF inequality of Eq.~\eqref{eq:lcineq} in~\cite{lugt22}.

Finally, although we saw that our approach can, in the SDI-QI setting, certify the causal nonseparability of the quantum switch even in the presence of substantial noise on the quantum switch, the NDI certification is less tolerant to noise on the entangled state shared between Phil and Emily.
Indeed, although we saw in Appendix~\ref{app:robustness} that the NDI certification is ``robust'', the actual reduction in Bell-inequality violation that can be tolerated in the self-testing procedure is typical small in absolute terms~\cite{bowles18a}.%
\footnote{We leave it as an open question to calculate the exact robustness, e.g.\ for the optimal witness inequality for the certification of the D-POVM obtained from the quantum switch from the operations of Eq.~\eqref{eq:instruqs}. Note, however, that with a noiseless maximally entangled state shared between Phil and Emily, the NDI certification can tolerate the same amount of noise on the quantum switch as the SDI-QI certification.}
The DRF inequality, on the other hand, appears to be much more robust to noise on the entangled state shared between Emily and Fiona after passing through the quantum switch, as any violation of the classical bound of $3/4$ of the last term in $\mathcal{I}_\text{DRF}$ (corresponding to a CHSH inequality between them) can potentially lead to a violation of the DRF inequality.
A more precise quantitive comparison of the robustness of the NDI and DRF certifications under different noise models would be an interesting direction of future study.

\end{document}